\documentclass[useAMS,usenatbib]{mn2e}

\usepackage[final]{graphicx}
\usepackage{wasysym}
\usepackage{amssymb}
\usepackage{subfigure}
\newcommand*{\myalign}[2]{\multicolumn{1}{#1}{#2}}

\title[New solar twins]{New solar twins and
  the metallicity and temperature scales of the Geneva Copenhagen Survey
  \thanks{Based on observations made with ESO
    Telescopes at the La Silla Observatory under programme ID 077.D-0525 and
    from the ESO Science Archive Facility under request number JDATSHCCCA119545 and following.}}

\author[Datson et al.]{Juliet Datson$^{1}$\thanks{E-mail: juliet.datson@utu.fi},
  Chris Flynn$^{2,3,4}$ and Laura Portinari$^{1}$\\ 
$^{1}$Tuorla Observatory, Department of Physics and Astronomy,
University of Turku, Finland,\\
$^{2}$Department of Physics and Astronomy, University of Sydney, NSW 2006 Australia\\
$^{3}$Finnish Centre for Astronomy with ESO, University of Turku, FI-21500, Piikkio, Finland\\
$^{4}$Centre for Astrophysics and Supercomputing, Swinburne University of Technology, VIC 3122 Australia}

\begin{document}

\date{Accepted 2012 July 16. Received 2012 July 2; in original form 2012 February 17}

\pagerange{\pageref{firstpage}--\pageref{lastpage}} \pubyear{2012}

\maketitle

\label{firstpage}

\begin{abstract}

We search for ``solar twins'' in the Geneva-Copenhagen Survey (GCS) using high
resolution optical spectroscopy. We initially select Sun--like stars
from the GCS by absolute magnitude, $(\mathrm{b}-\mathrm{y})$ colour and
metallicity close to the solar values. Our aim is to find the stars which
are spectroscopically very close to the Sun using line depth ratios and the
median equivalent widths and depths of selected lines with a range of excitation potentials. We
present the ten best stars fulfilling combined photometric and spectroscopic
criteria, of which six are new twins. 

We use our full sample of Sun--like
stars to examine the calibration of the metallicity and temperature scale in
the GCS. Our results give rise to the conclusion that the GCS
may be offset from the solar temperature and metallicity for sun-like stars by
100\,K and 0.1\,dex, respectively.

\end{abstract}

\begin{keywords}
stars: abundances -- stars: fundamental parameters -- stars: solar-type
\end{keywords}

\section{Introduction}

``Solar twins'' (or ``solar analogues'') are stars which are very close matches
to the spectroscopic and photometric properties of the Sun \citep{b6}, and
while there is currently no ``perfect twin'', a few stars are
known which are very close matches to the Sun. For over a decade, the star
considered most similar to the Sun has been 18\,Sco/HD\,146233 \citep{b2,b1},
and recent asteroseismological and interferometric measurements have confirmed
its radius and mass to be solar within a few percent \citep{b33}.  Based on its
spectroscopic properties, HD\,98618 was considered the next best solar twin by
\citet{b3}, while \citet{b13} have found HIP\,100963 to be as good a twin as
18\,Sco (although both stars have a higher Li abundance than the Sun by 0.5 and
0.8\,dex respectively). Currently, HD\,56948 is the star considered closest to
the Sun \citep{b11} -- its lithium abundance is very similar to the Sun,
and together with another close twin (HIP\,73815), it shows that the solar lithium
abundance is not atypical. Some tens of solar twins (or solar analogues)
have been published in recent years \citep{b13,b32}; and a solar twin has even
been identified in the open cluster M67 \citep{b5}. To date all solar twins
show small but interesting differences with respect to the Sun: such as the
lithium abundance being high \citep{b3} or the stars being variable and showing
chromospheric activity (e.g. 18\,Sco, \citealt{b4}).

Solar twins are useful because, obviously enough, one cannot point the same
instruments/telescopes at the Sun as used on faint objects in the night
sky. This gives rise to a major difficulty with calibrating the stellar
metallicity and temperature scale to the same scale as the Sun, in which these
parameters are measured (notwithstanding that the absolute solar
metallicity has been under considerable discussion in the last decade,
\citealt{b20}).

The largest extant sample of solar type stars (F, G and K dwarfs), the
Geneva-Copenhagen-Survey (hereafter GCS, \citealt{b9}), has a photometric
metallicity and temperature scale which is tied to the stellar
colours. Recently, the metallicity and temperature calibrations have been
called into question by \citet{b24}, who used the InfraRed Flux Method
on 423 stars, and the properties of 10 solar twins, to argue that the GCS scale
may need shifting by about 100\,K in temperature and 0.1\,dex in metallicity.  We
address this topic in this paper, presenting a new method for testing the
scales in the GCS catalogue using solar analogues, and finding a similar shift.

Interest has revived in solar twins since the advent of exoplanet detection.
Interestingly, the first discovered exoplanet host, 51 Peg
\citep{b7}, is a very good solar twin \citep{b6}. Exoplanets have been found around
other solar twins \citep{b28,b29}, but to date no systematic searches have been
undertaken. One of our aims is to provide a list of nearby solar twins by
searching systematically for them in the GCS.

Recently \citet{b32} and \citet{b15} have suggested that whether a star hosts
exoplanets or not might be revealed in the detailed chemical composition
obtained from high resolution spectra. Another motivation for this study is to
provide new targets to probe this correlation further.

In this paper we present the results of our own quest for solar twins, based on
photometric selection from the GCS catalogue, combined with
high resolution spectroscopic data. Most previous surveys looking for solar
twins have focused on objects of the Northern Hemisphere, such as at the
Observatoire des Haute Provence \citep{b1}, Keck \citep{b3,b8} or the McDonald
Observatory in Texas \citep{b15}. We explore the relatively understudied
Southern Hemisphere from the Max-Planck-Gesellschaft (MPG)/European Southern
Observatory (ESO) 2.2m at La Silla Observatory, giving us the
opportunity to extend our search to targets not considered previously, and we
turn up with six new twins.

In this paper we provide a description of our photometric candidate selection
process in section 2; in section 3 we outline the observations and data
reduction; in section 4 we present our methods for finding solar twins and the
results.  In section 5 we use our full spectroscopic sample of about a hundred
Sun--like stars to test the temperature and metallicity scale in the GCS, by
differential comparison to the solar spectrum, and we draw our conclusions in
section 6.

\section[]{Candidate selection}

Our candidate solar-twins were selected from the first release of the
Geneva-Copenhagen-Survey (GCS-I) \citep{b9} of Stromgren colours, absolute
magnitudes, metallicity and temperature estimates for $\sim14\,000$ nearby F to
K type stars. We selected stars bracketting the solar $(\mathrm{b}-\mathrm{y})$ colour, absolute
visual magnitude $\mathrm{M}_\mathrm{V}$ and metallicity, for which we adopt the solar values of
$(\mathrm{b}-\mathrm{y})_\odot = 0.403$ \citep{b19}, $\mathrm{M}_\mathrm{V} = 4.83$ \citep{b22} and [Fe/H] $=0.0$
(by definition).  Our ranges are : $0.371 < (\mathrm{b}-\mathrm{y}) < 0.435$, in absolute
magnitude $4.63 < \mathrm{M}_\mathrm{V} < 5.03$ and in metallicity (GCS-I scale) $-0.15 <
[\mathrm{Fe/H}] < 0.15$. These criteria resulted in 338 stars, of which 80 were
chosen for the proposal as accessible from La Silla, and not already available
in the Fiber-fed Extended Range Optical Spectrograph (FEROS) archive. In the end, 
70 of these 80 targets were observed (all in service mode).

\begin{figure}
	\includegraphics[width=90mm]{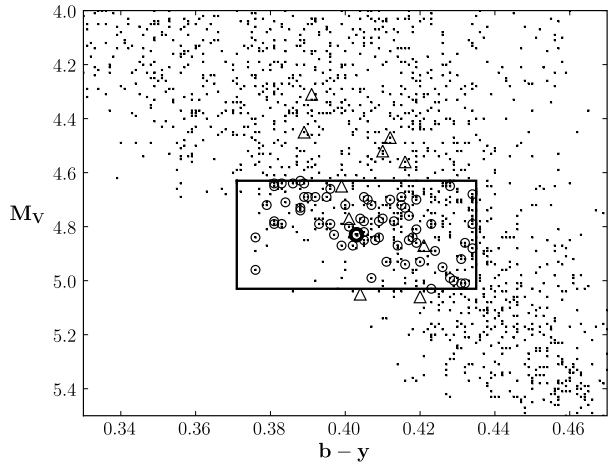}
	\caption{Stars from the Nordstr\"om et al. (2004) catalogue (GCS-I)
          (dots), in the metallicity range of $-0.15 < [\mathrm{Fe/H}] < 0.15$,
          together with our initial twin candidates (circles) and the
          candidates from \citet{b1}(triangles). The Sun is marked in the
          middle of the box. The solar twin candidates of the \citet{b1} sample
          (triangles) tend to lie redward (cooler) and tend to be
          intrinsically brighter than the Sun.}
	\label{proposal}
\end{figure}

In Fig.\,\ref{proposal} we show the colour-magnitude diagram (CMD) of the main
sequence and turnoff stars from the GCS-I catalogue. The dots show our initial
photometric twin candidates, chosen for spectroscopic follow--up.  An
interesting effect appears if we show the 10 very good spectroscopically
matched solar twins found by \citet{b1} (triangles). Their twins tend to
lie redward (cooler) and tend to be brighter than the Sun. This highlights the
point (concluded by those authors) that purely spectroscopic matching might 
still yield stars with systematic differences to the Sun in other properties --  thus, in this paper, we examine
  the effects of using both spectroscopic and photometric criteria to find
  solar twins.

While this project was being undertaken, two revisions of the Geneva-Copenhagen
Survey (GCS-II and GCS-III) \citep{b27, b21} were published. Both revisions
addressed the metallicity, age and temperature scales of the stars, and
utilised the updated Hipparcos parallaxes \citep{b26} when they became
available. This resulted in increased temperatures, which went up by an average
of 80\,K at solar values and in decreased metallicities, which went down by an average
of 0.05\,dex.

With the revised absolute magnitudes and metallicities, some of our target
stars moved slightly out of the original selection window while others
moved in. These new arrivals have been included in our sample, as we searched
the FEROS archive at ESO and found spectra for 28 of these candidates. We also included
stars in our sample which have been classified as 'Sun-like' stars by
others \citep{b12,b13,b14,b15,b1}. This yielded another 47 stars, bringing our
total sample to 145 objects. They range in apparent $V$ magnitudes from 3.5 to
9, with the majority lying around $V = 8$.

\begin{figure}
	\includegraphics[width=90mm]{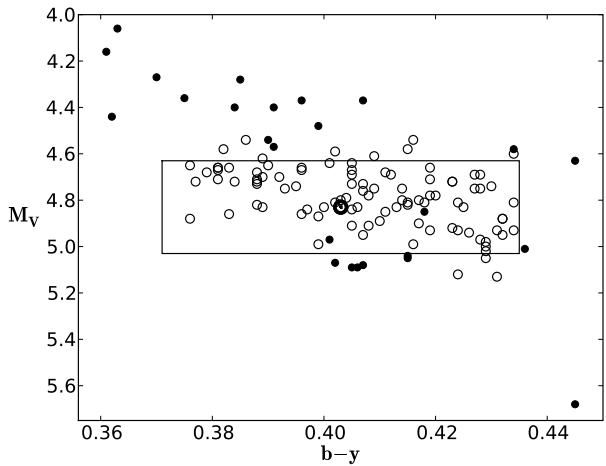}
	\caption{As Fig.\,\ref{proposal}, showing our target sample as open
          circles (from GCS-I and GCS-III) and the targets taken from other
          papers (filled circles). The Sun is shown in the middle of the box,
          which represents our original selection window.}
	\label{newsample}
\end{figure}

Fig.\,\ref{newsample} shows our final spectroscopic sample in the $\mathrm{M}_{\mathrm{V}}$
vs. $(\mathrm{b}-\mathrm{y})$ plane. Our basic candidate stars, initially selected from the GCS-I,
but with revised GCS-III data and stars which entered the selection window as a
result of the revised GCS data, are shown as open circles. Finally, Sun-like
stars from other studies in the literature are shown as filled circles. The
Sun's location is shown in the middle of our selection box.

Some stars selected from the literature lie very far in absolute magnitude and
colour from our selection window. This is because often the authors adopted a
broad definition of Sun--like stars, sometimes extending to F and K dwarfs
\citep{b25} or metal-rich stars \citep{b12}.  We decided to keep all those
stars in our analysis, to have a chance to test broad trends with temperature
and metallicity in the spectroscopic selection of our best twin candidates.

\section{Observations}

Spectra of 70 of our 80 target stars were obtained with the FEROS instrument
\citep{b36} on the MPG/ESO 2.2 meter telescope at La Silla, Chile. FEROS is an
echelle spectrograph with a resolution of $R \sim 48000$ and covers a
spectral range of 3500-9200\,\AA, permitting the analysis of a wide range of
lines. Spectra were obtained in service mode over the period 77A, in July and
August 2006. Exposure times varied from 200\,s up to 950\,s (depending on object
magnitude), resulting in typical signal to noise (S/N) ratios of $\sim$100-150.

We also obtained spectra from the FEROS archive if they had exactly the same
instrumental setup as our service observations, resulting in an additional 75
stars. These spectra were taken between 2003 to 2008. We have checked, by means of
twilight spectra and repeated observations of $\tau$ Ceti from the archive, and
found that the instrument is very stable over these long time scales (details are in 
section 4.5). The S/N for these additional spectra are also
of the order of 100-150. High S/N spectra of the asteroid Ceres were also
taken as our solar spectrum comparison.

Data reduction was done with the FEROS pipeline, resulting in 1-D spectra that
were flatfielded, bias subtracted, sky subtracted and wavelength
  calibrated in the range of 3500-9200\,\AA. The wavelength calibration is based on
ThAr+Ne arcs and the pipeline rebins the spectra to a linear
0.03\,\AA\ resolution over their full wavelength range.

The final FEROS pipelined spectra contained a number of significant wiggles in
the continuum level over short and long scales. These were far too complex
  to remove using polynomial or spline fits over the whole spectral range. We
took the simple approach of fitting piecewise 10\,\AA\ sections of the spectrum,
and normalising to the mode of the histogram of pixel values in each
section. This procedure produced very flat spectra in regions with few or weak
lines (i.e. most of the spectrum). All the lines analysed were in regions with
only a few weak lines, and this simple procedure was found to be entirely
adequate. Stars for which we had several spectra showed that this procedure was also
very stable.

During the reduction process we found some spectra to be too noisy to be
flattened properly and some spectra with unaccountable continuum jumps. These
spectra were discarded, reducing our sample to 100 stars.


\section{Analysis}

A number of approaches have been adopted to finding solar twins by groups
active over the last decade. These are : $\chi^2$ matching over a large range
in the spectral energy distribution (SED) \citep{b1}; comparison of the
equivalent widths of specific iron lines relative to the Sun, \citep{b25, b11};
comparison of the line depth of specific iron lines \citep{b3} or line-by-line differential
abundance comparison \citep{b5}. In this paper we apply four methods to our
sample with the view that good twins should stand out in more than one.

Specifically, our methods are comparisons of:

\begin{enumerate}

  \item the median differences between the equivalent widths (EWs)
    in the target stars and in Ceres, for 109 lines covering a range of species
    and ionisation states (Section 4.2), which is closely related to the 
    ``first criterion'' method of \cite{b3} 
    
  \item the relative differences between the EWs of 33 FeI lines in the star and
    in Ceres, as a function of the excitation potential of the lines (Section 4.3),
    a method similar to that used by \cite{b11}

  \item the relative differences in FeI line depths of the same 33 lines, as a function of the
    excitation potential of the lines (Section 4.4) (the same method as in \citealt{b3})
    
  \item line depth ratios for specific pairs of high and low excitation potential 
    lines (Section 4.5), a traditional method to derive temperatures (see e.g. \cite{b16})

\end{enumerate}

Method (i) makes use of the equivalent widths of 109 lines for 20 elements
(kindly provided to us by Ivan Ram{\'{\i}}rez --- private comm.). The lines
cover the spectral range of 5000-8000\,\AA\ and are carefully selected to be
unblended, weak and without telluric contamination. The species measured are
OI, NaI, MgI, AlI, SiI, SI, KI, CaI, ScII, TiI, VI, CrI, MnI, FeI+II, CoI, NiI,
CuI, ZnI, ZrII and BaII. The median depths of these lines relative to their
depths in the comparison solar (Ceres) spectrum are used to search for solar
twins.

In Method (ii) and (iii) we confined the twin matching process to the EWs and
LDs of just 33 FeI lines \citep{b3}. Temperature and metallicity sensitivity
in these lines is attained by comparing the median depths of the lines relative
to the lines in the Ceres spectrum, as a function of their excitation
potential.

Method (iv) uses the line depth (LD) ratios of three pairs of lines very close
in wavelength, with one line having a low and one a high excitation
potential. This is a classical method to probe temperature, as the line ratios
are known to have negligible metallicity sensitivity (as we confirm in section
4.5).

\subsection{Measuring equivalent widths and TWOSPEC}

We developed a code, TWOSPEC, to measure the equivalent widths of the selected
lines. The program compares two spectra at the same time -- the target star
being analysed and the spectrum of Ceres as the comparison. Equivalent widths
are measured simply by computing the missing light in the line in a window
300\,m\AA\ wide, relative to the continuum level in each spectrum. The
placement of the two continua in the spectra is an important issue, since we
are doing differential spectroscopy between the star and Ceres, thus if the two
continua are systematically different, there will also be a systematic
difference in the measured equivalent widths. Some well known methods for
setting the continuum are to define good continuum regions around each line, or
to search for the mode of flux values in the selected window. We choose a third
way --- we normalise the two spectra by the total flux in each in a
10\,\AA\ window around the target line. We achieve two advantages in doing this
-- since we are searching for stars as similar to the Sun as possible -- a
perfectly matching spectrum between both the star and Ceres in this window will
have the same missing light due to the same lines, and the continuum will be
accurately set. For stars which are a slight mismatch to the Sun, the continuum
will be commensurately offset, and as this will tend to emphasise the mismatch
with the Sun, this aids us in finding good spectroscopic matches. Tests of this
procedure show that it works very well. For example, when setting the continumm
using the flux normalisation technique, and comparing spectra of the same star
HD\,147513 to Ceres, yielded a scatter in the measured EWs of 95 lines of
1.4\,m\AA (compared to their EWs in Ceres), whereas our best effort to fit the
continuum directly resulted in a scatter in the measured EWs for the same 95
lines of 1.9\,m\AA\ relative to Ceres.

TWOSPEC displays to the user both the target and reference spectra centred on
the target line, the assigned continuum level, and the window around the line
used to measure the EW. The depths of the lines were also measured by
estimating the centre of light by fitting a parabola to the three lowest points
in the line. We tried Gaussian fitting of the lines to better measure the EWs,
but this did not lead to any significant improvement in the EWs -- as measured
by comparing multiple spectra of the same object -- so this technique was not
utilised. All lines in all the spectra were inspected by eye by two of us (JD
and CF) in this way, allowing us to develop a good impression of the quality of
the match for each star, as well as enabling us to drop bad spectra or bad
lines due to poor flat fielding, electronic readout issues or cosmic
rays. TWOSPEC requires no user input beyond viewing the lines individually and
confirming that the data are good. Once a few bad spectra had been spotted and
dropped, the complete data set could be reduced in batch mode with no user
intervention. This made experiments, such as how we set the continuum or
measured the EWs or LDs, very straightforward to test.

We also measured EWs by hand for 10 random stars using IRAF. The IRAF and
TWOSPEC EWs were found to be in very good agreement, with a scatter of a few
m\AA, and no significant systematic offsets.

\begin{figure}
	\includegraphics[width=90mm]{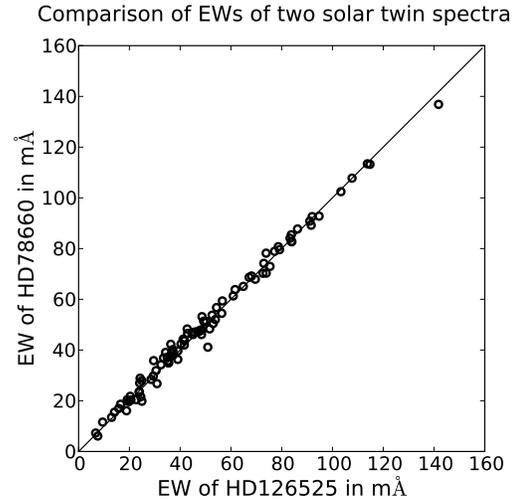}
	\caption{A comparison of measured equivalent widths for the two stars
          HD\,126525 and HD\,78660. The line is the 1:1 relation.
        }
	\label{EWs}
\end{figure}

Fig.\,\ref{EWs} shows a comparison of the EWs for the 109 lines of two of our
stars, HD\,126525 and HD\,78660. The lines range in EW from $\sim$10\,m\AA\ to 
140\,m\AA, with most lines being in the range of 20-80\,m\AA.

\subsection{Method (i): comparison of EWs for a range of species}

This method is closely related to the ``first criterion'' method of
\citet{b3}. We measured the median $\textless\Delta
\mathrm{EW}_{\mathrm{all}}\textgreater$ and scatter $\chi^{2}(\Delta
\mathrm{EW}_{\mathrm{all}})$ of the differences $\Delta$EW$_{\mathrm{all}}$ in
the EWs of target stars relative to Ceres, for the $\mathrm{N}=109$ lines:

\begin{equation}
  \Delta \mathrm{EW}_{\mathrm{all}} = (\mathrm{EW}(\star)-\mathrm{EW}(\astrosun))/\mathrm{EW}(\astrosun),
\end{equation}

\begin{equation}
  \chi^{2}(\Delta \mathrm{EW}_{\mathrm{all}}) = \sum_{i=1..\mathrm{N}}
  ((\mathrm{EW}_{i}(\star)-\mathrm{EW}_{i}(\astrosun))/\mathrm{EW}_{i}(\astrosun))^{2},
\end{equation}

\noindent where the solar values refer to the Ceres spectrum. 

\begin{figure*}
\centering
\subfigure{
  \includegraphics[scale=0.35]{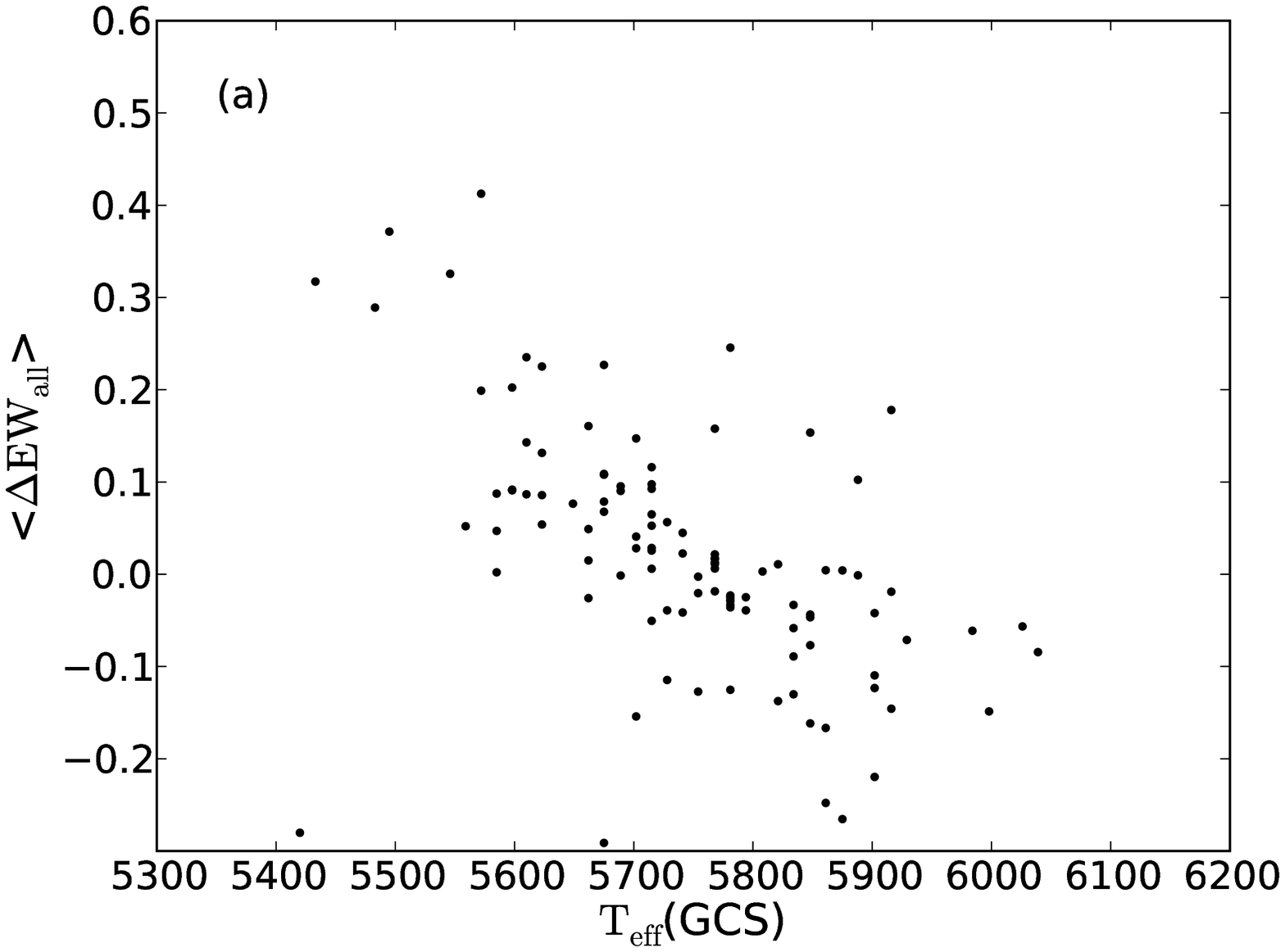}
\label{fig:subfig1}
}
\subfigure{
  \includegraphics[scale=0.35]{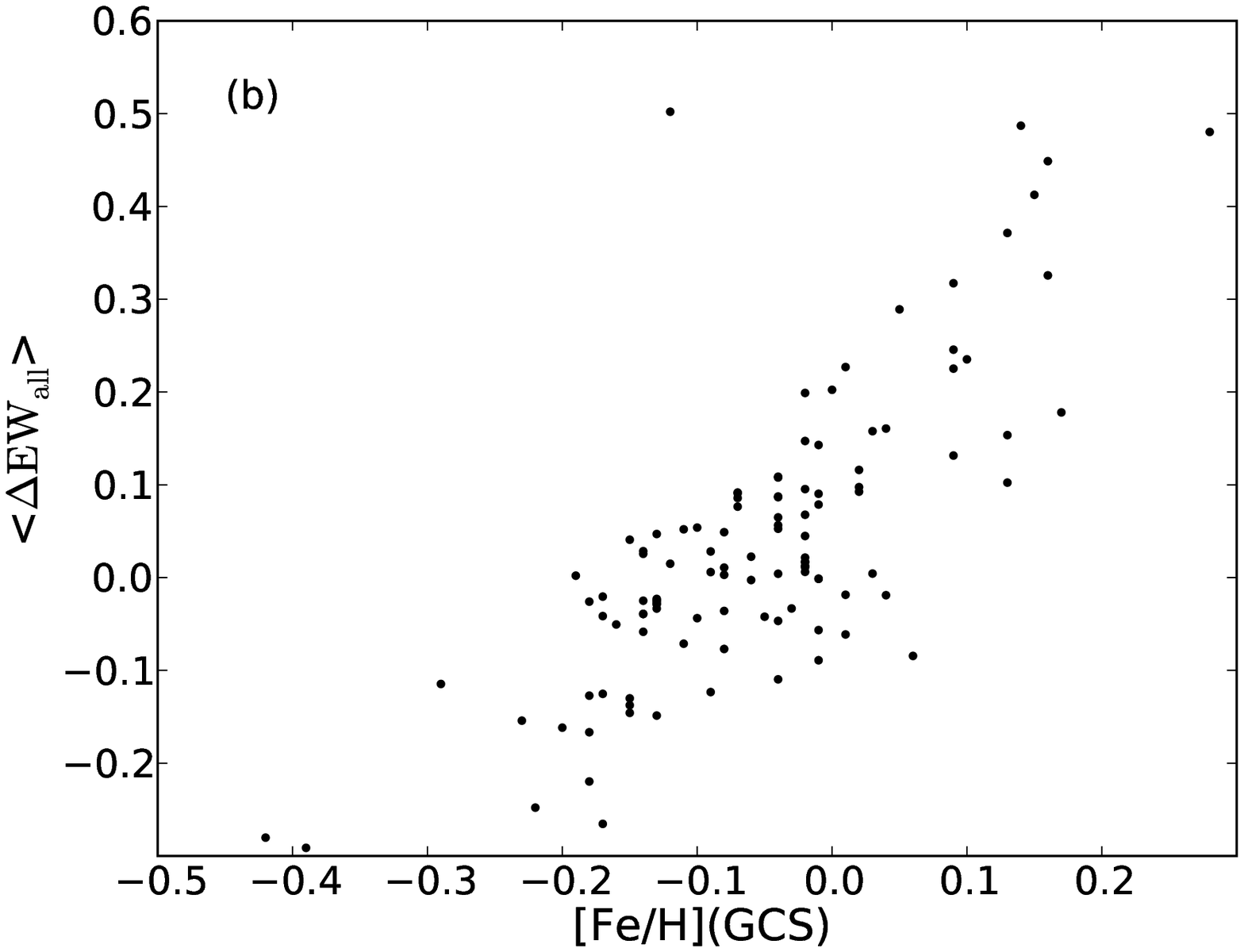}
\label{fig:subfig2}
}
\subfigure{
  \includegraphics[scale=0.35]{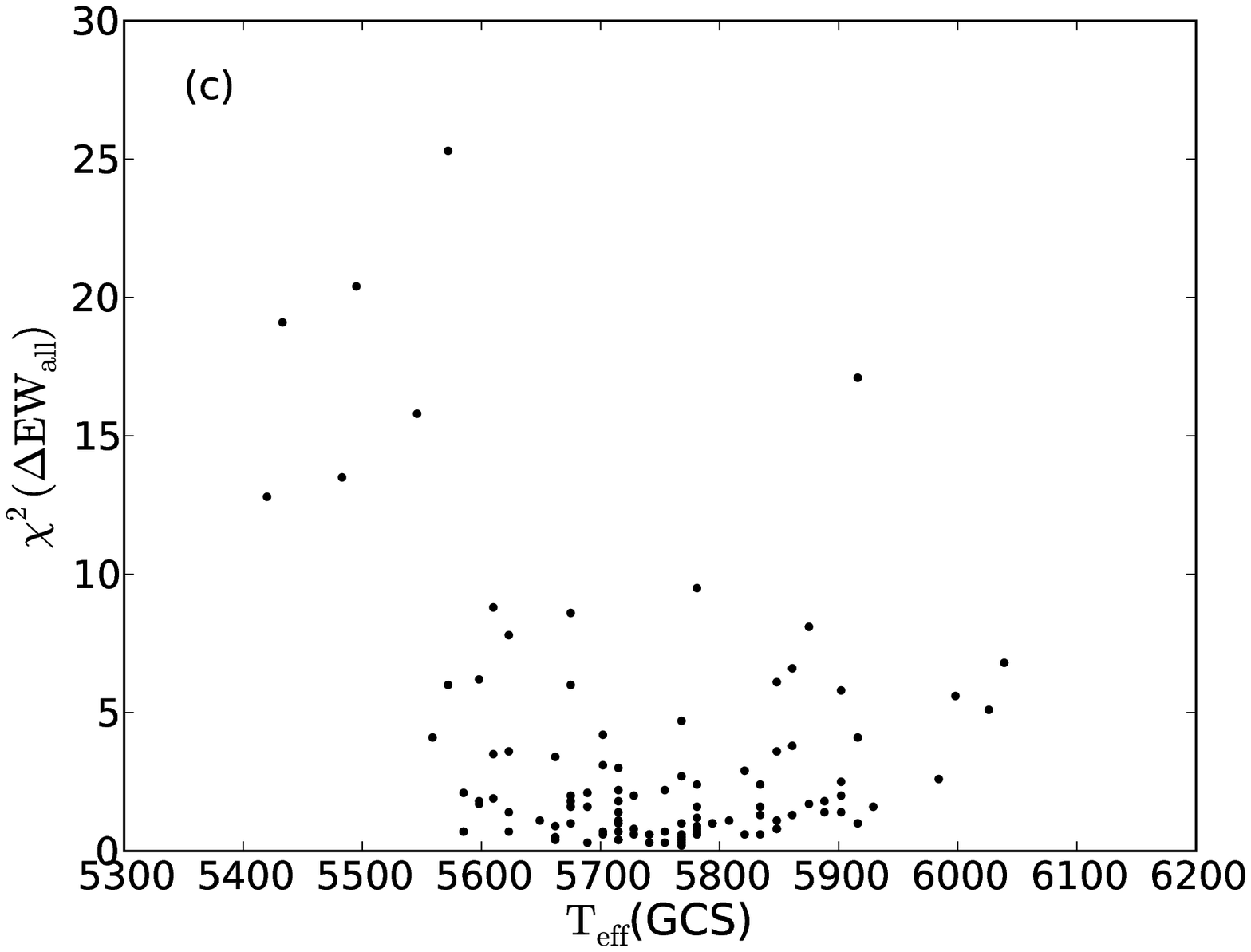}
\label{fig:subfig3}
}
\subfigure{
  \includegraphics[scale=0.35]{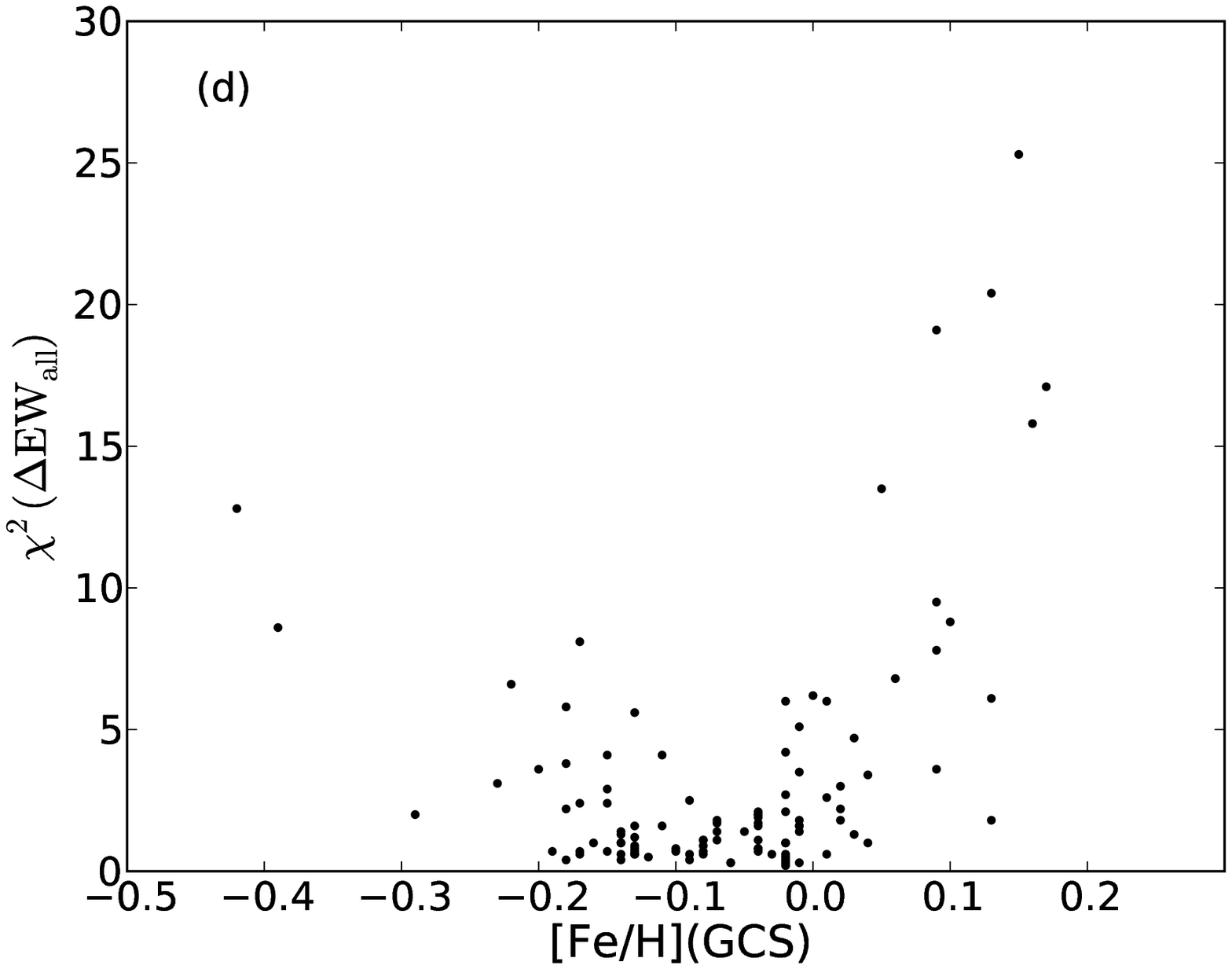}
\label{fig:subfig4}
}
\caption[Optional caption for list of figures]{Method (i). Panel (a): Median
  relative difference of 109 lines of various species in our 100 target stars
  in comparison to Ceres, plotted as a function of the effective temperature
  from GCS-III.  $\textless\Delta\mathrm{EW}_{\mathrm{all}}\textgreater$ is
  anti-correlated with temperature. Panel (b): Same as panel (a), but plotted
  against the metallicities of the stars from GCS-III.
  $\textless\Delta\mathrm{EW}_{\mathrm{all}}\textgreater$ is correlated with
  [Fe/H], as one would expect. Panel (c): Scatter of the relative EWs of the
  stars compared to Ceres, shown as a function of temperature from GCS-III. The
  scatter in the relative EWs of the lines increases very rapidly as one moves
  to cooler or hotter stars relative to the Sun at 5777\,K. Panel (d) as for
  panel (c), but plotted versus the stellar metallicities. The scatter
  increases rapidly as the stellar metallicities depart from solar in either
  direction. Note that one can see here already that the scatter is least at a
  metallicity below the solar value, indicating that the GCS-III metallicity
  scale might be too metal-poor for solar type stars (as analysed in depth
  section 5). (We have excluded four stars with high $\chi^{2}(\Delta
  \mathrm{EW}_{\mathrm{all}})$ for clarity).}

\label{fig:chi2}
\end{figure*}

The results for our 100 stars samples are shown in Fig.\,\ref{fig:chi2},
plotted as functions of the (GCS-III) temperature and metallicity.

For a solar twin, the median difference $\textless\Delta
\mathrm{EW}_{\mathrm{all}}\textgreater$ vanishes (to within observational
error) and $\chi^{2}(\Delta \mathrm{EW}_{\mathrm{all}})$ should be consistent
with observational scatter alone. We define solar twins in our method (i) as
having $\chi^{2}(\Delta \mathrm{EW}_{\mathrm{all}}) \leq 1$ and a
$\textless\Delta \mathrm{EW}_{\mathrm{all}}\textgreater$ = 0 within 2$\sigma$.
This results in 7 solar twins, and these are listed in Table 1.

\begin{table}
 \centering
     \caption{List of solar twins using $\chi^{2}(\Delta
       \mathrm{EW}_{\mathrm{all}})$ (i.e. method (i)), ordered by $\chi^{2}(\Delta
       \mathrm{EW}_{\mathrm{all}})$ (see also Figure \ref{fig:chi2}).}
  \begin{tabular}{@{}lcr@{}}
  \hline
   Name & $\chi^{2}(\Delta \mathrm{EW}_{\mathrm{all}})$ 
        & \myalign{c}{$\textless\Delta \mathrm{EW}_{\mathrm{all}}\textgreater$}\\
 \hline
 HD\,146233 & $0.2\pm0.4$ & $ 0.006\pm0.005$ \\
 HD\,97356     & $0.3\pm0.4$ & $-0.003\pm0.006$ \\
 HD\,138573    & $0.3\pm0.4$ & $-0.001\pm0.006$ \\
 HD\,78660     & $0.4\pm0.4$ & $ 0.006\pm0.007$ \\
 HD\,117860    & $0.6\pm0.4$ & $ 0.011\pm0.008$ \\
 HD\,126525    & $0.7\pm0.4$ & $ 0.002\pm0.009$ \\
 HD\,142415    & $1.0\pm0.4$ & $-0.019\pm0.010$ \\
\hline
\end{tabular}
\end{table}

\subsection{Method (ii): comparison of FeI equivalent widths versus excitation potential}

Method (ii) is based on the technique used by \citet{b11}. We used only the 33
Fe I lines (from our total list of 109 lines), and computed a new
$\textless\Delta \mathrm{EW}_{\mathrm{FeI}}\textgreater$ as the median of:

\begin{equation}
  \Delta \mathrm{EW}_{\mathrm{FeI}} =
  (\mathrm{EW}_{\mathrm{FeI}}(\star)-\mathrm{EW}_{\mathrm{FeI}}
  (\astrosun))/\mathrm{EW}_{\mathrm{FeI}}(\astrosun)
\end{equation}

\begin{figure}
	\includegraphics[width=90mm]{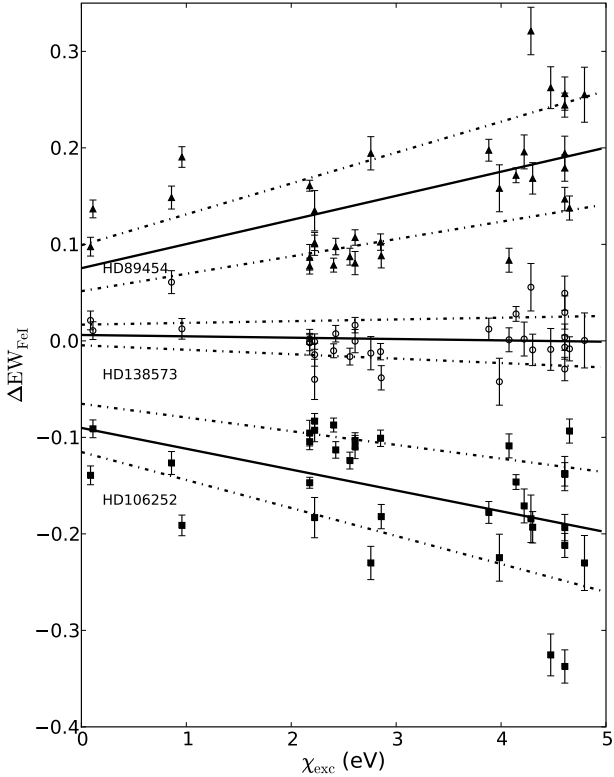}
	\caption{Method (ii). The relative difference in EW of 33 Fe I lines
          compared to Ceres, plotted as a function of the excitation potential
          of the lines for three example stars. The solid line shows the least
          square fit, and the dotted lines are the 1$\sigma$ error limits, for
          each star. HD\,138573 is an excellent solar twin, having, within
          observational error, no median difference in the EW of the lines
          relative to Ceres and no dependence on the excitation potential. For
          comparison HD\,89454 and HD\,106252 are poor solar twins, as the EWs
          of the lines show a clear dependence on excitation potential.  The
          relative difference in EWs and the slope[($\Delta\mathrm{EW}_{\mathrm{FeI}}$) 
          vs. $\chi_{\mathrm{exc}}$]
          are thus good proxies for the metallicity and temperature of the star
          relative to the Sun (as extensively discussed by \citealt{b3} and
          \citealt{b11}).  }
	\label{HD78660}
\end{figure}

We also measure the slope of the relation between $\Delta
\mathrm{EW}_{\mathrm{FeI}}$ and the excitation potential
($\chi_{\mathrm{exc}}$) of each Fe I line (see Fig.\,\ref{HD78660}). 

\begin{figure}
	\includegraphics[width=90mm]{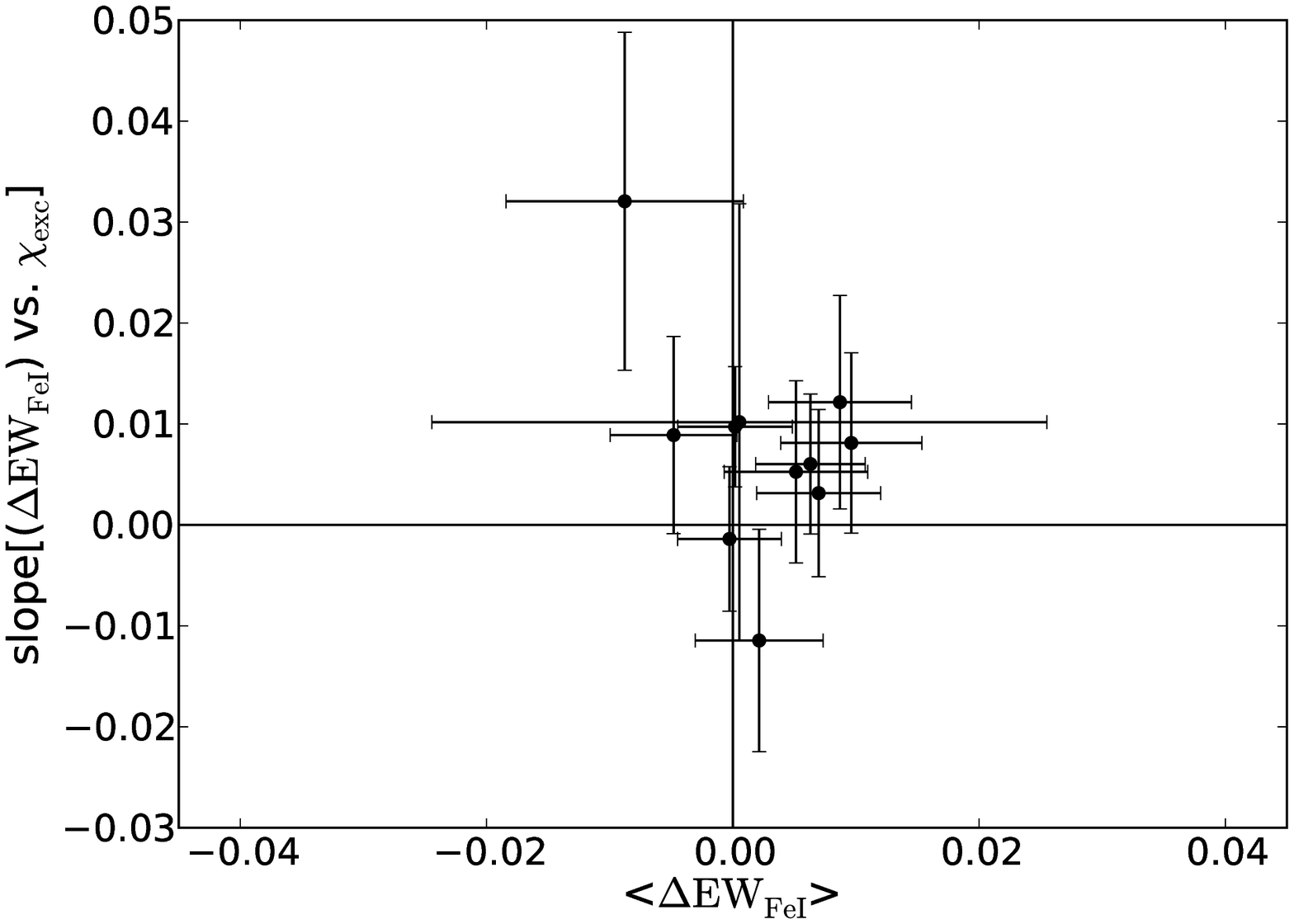}
	\caption{A close-up of the best targets for method (ii). These stars have $\textless\Delta
          \mathrm{EW}_{\mathrm{FeI}}\textgreater$ and slope[($\Delta
          \mathrm{EW}_{\mathrm{FeI}}$) vs. $\chi_{\mathrm{exc}}$] of 0 within $2
          \sigma$, here plotted with errorbars of $1\sigma$ for clarity.}
\label{slopemedianEW}
\end{figure}

Figure \ref{slopemedianEW} shows stars which are good matches to the Sun, where
we consider good matches to be stars for which $\textless\Delta
\mathrm{EW}_{\mathrm{FeI}}\textgreater$ and a slope[($\Delta
\mathrm{EW}_{\mathrm{FeI}}$) vs. $\chi_{\mathrm{exc}}$] vanishes to within
$2\sigma$, where $\sigma$ is the observational error.

Eight targets satisfy these criteria, and are shown in Table 2 and
Fig.\,\ref{slopemedianEW}. Five of these targets are common with the stars found in
method (i) -- they are HD\,78660, HD\,117860, HD\,126525, HD\,138573 and HD\,146233.

\begin{table}
\centering
\caption{List of solar twins from method (ii), ordered by name (see also Figure \ref{slopemedianEW}).}
\begin{tabular}{@{}lrr@{}}
\hline
Name & \myalign{c}{$\textless\Delta \mathrm{EW}_{\mathrm{FeI}}\textgreater$} 
     & \myalign{c}{slope[($\Delta \mathrm{EW}_{\mathrm{FeI}}$) vs. $\chi_{\mathrm{exc}}$]}\\
\hline
HD\,78660  & $0.007\pm0.005$ & $0.003\pm0.008$\\
HD\,117860 & $0.009\pm0.006$ & $0.012\pm0.011$\\
HD\,126525 & $0.002\pm0.005$ & $-0.011\pm0.011$\\
HD\,138573 & $0.000\pm0.004$ & $-0.001\pm0.007$\\
HD\,146233 & $0.000\pm0.005$ & $0.010\pm0.006$\\
HD\,147513 & $-0.005\pm0.005$ & $0.009\pm0.010$\\
HD\,163441 & $0.005\pm0.006$ & $0.005\pm0.009$\\
HD\,173071 & $-0.009\pm0.010$ & $0.032\pm0.017$\\
\hline
\end{tabular}
\end{table}

\subsection{Method (iii): comparison of FeI line depths versus excitation potential}

In method (iii) we used the same technique as method (ii), but computed the
$\textless\Delta \mathrm{LD}_{\mathrm{FeI}}\textgreater$ (line depth) for the
iron lines rather than the EWs (i.e. the exact same technique as in \citealt{b3}):

\begin{equation}
  \Delta \mathrm{LD}_{\mathrm{FeI}} =
  (\mathrm{LD}_{\mathrm{FeI}}(\star)-\mathrm{LD}_{\mathrm{FeI}}(\astrosun))/
  \mathrm{LD}_{\mathrm{FeI}}(\astrosun)
\end{equation}

\begin{figure}
	\includegraphics[width=90mm]{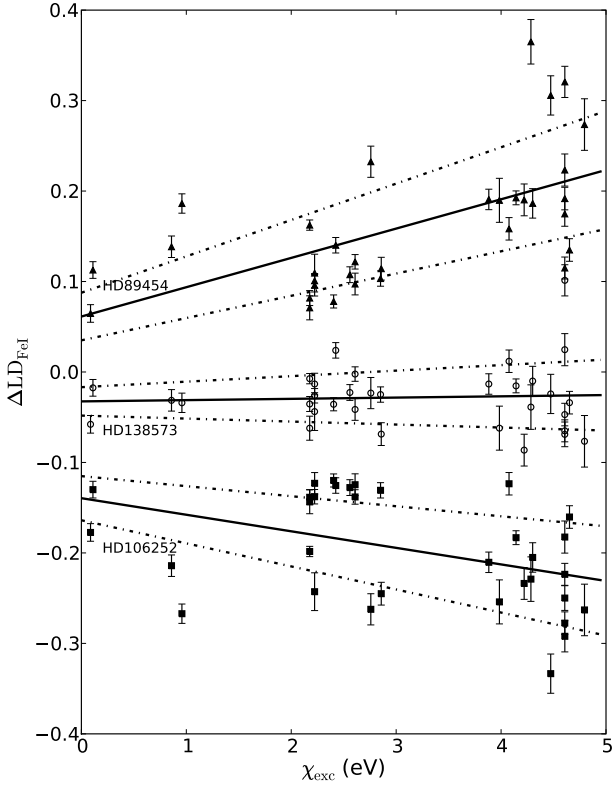}
	\caption{Method (iii) uses the depths of 33 FeI lines in the target
          star relative to the lines in the Ceres spectrum, and the excitation
          potential of the lines, to search for solar twins.  We show here the
          the relative difference between the star and Ceres for all 33 FeI
          lines as a function of excitation potential, for three stars:
          HD\,138573, a very close match to the Sun according to method (iii);
          and two poor matches to the Sun (HD\,89454 and HD\,106252).}
    	  \label{HD138573}
\end{figure}

The slope of the relative line depth differences as a function of the
excitation potential of the lines was determined as with method (ii)
(Fig.\,\ref{HD138573}).

\begin{figure}
	\includegraphics[width=90mm]{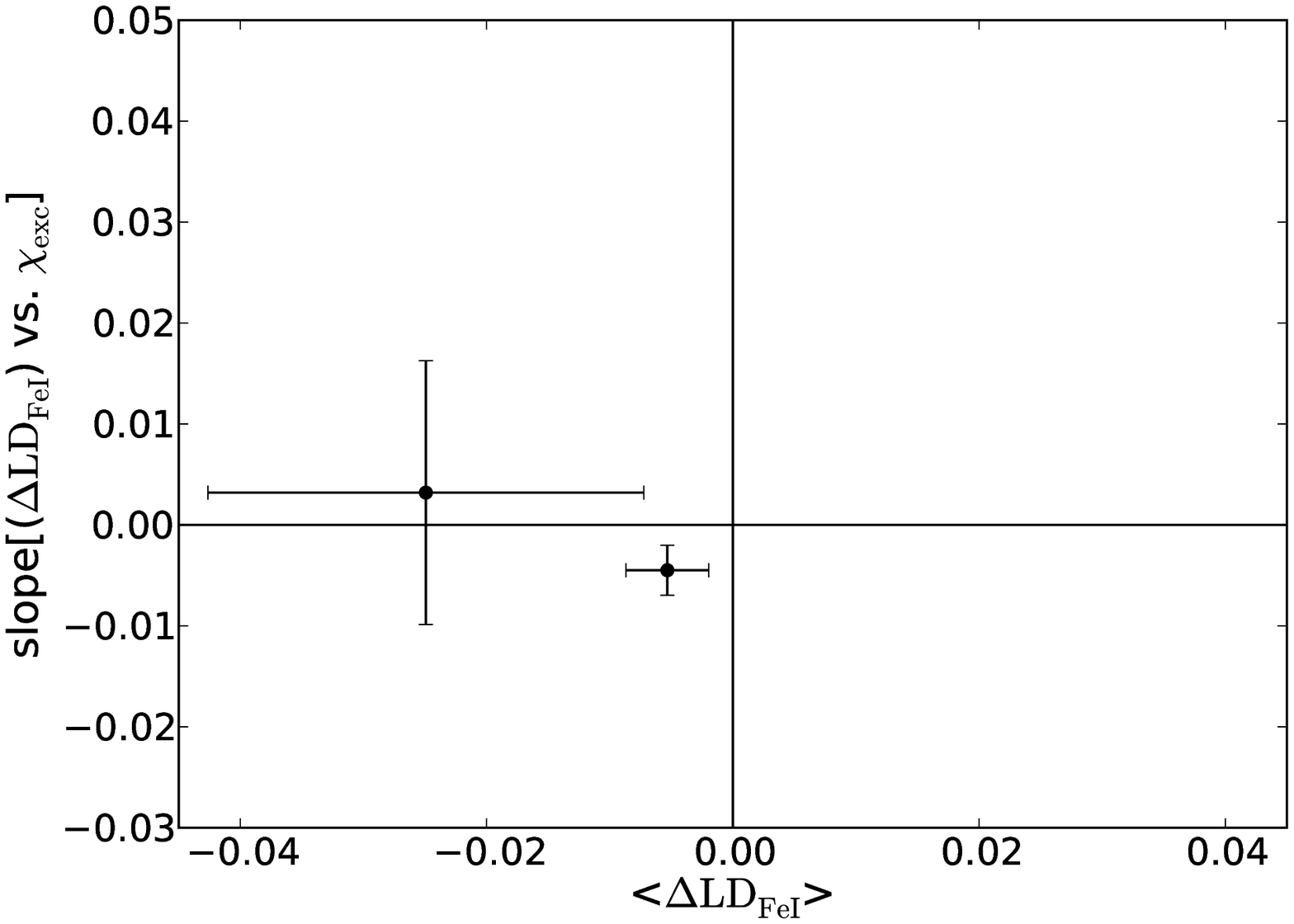}
	\caption{A close-up of the best targets using method (iii). These stars have median
          $\textless\Delta \mathrm{LD}_{\mathrm{FeI}}\textgreater$ and
          slope[($\Delta \mathrm{LD}_{\mathrm{FeI}}$) vs. $\chi{\mathrm{exc}}$]
          of 0 within $2\sigma$, here plotted with errorbars of $1\sigma$ for
          clarity.}
	\label{slopemedianiron}
\end{figure}

Our most Sun-like targets are shown in Fig.\,\ref{slopemedianiron} and are
listed Table 3. They are HD\,126525 and HD\,146233, as they both have indices
which vanish to within $2\sigma$. Both stars turn up also in the previous two
lists of solar twins.

\begin{table}
 \centering
     \caption{List of solar twins from method (iii), see also Figure \ref{slopemedianiron}.}
  \begin{tabular}{@{}lrr@{}}
  \hline
Name & \myalign{c}{$\textless\Delta \mathrm{LD}_{\mathrm{FeI}}\textgreater$} 
     & \myalign{c}{slope[($\Delta \mathrm{LD}_{\mathrm{FeI}}$) vs. $\chi{\mathrm{exc}}$]}\\
 \hline
HD\,126525 & $-0.005\pm0.003$ & $-0.004\pm0.002$\\
HD\,146233 & $-0.025\pm0.018$ & $0.003\pm0.013$\\
\hline
\end{tabular}
\end{table}

\subsection{Method (iv): Line depth ratios for choice line pairs}

This method is based on the technique of \citet{b16}, who use line depth ratios
(LDR) for carefully choosen pairs of lines very close in wavelength to probe
excitation temperatures. We used the following pairs: FeI (6089.5\,\AA)/VI
(6090.2\,\AA); VI (6243.1\,\AA)/SiI (6243.8\,\AA) and VI (6251.8\,\AA)/FeI
(6252.6\,\AA).

We compare the relative difference of these line ratios in the target star and
in Ceres, as follows:

\begin{equation}
  \Delta \mathrm{LDR} =
  (\mathrm{LDR}(\star)-\mathrm{LDR}(\astrosun))/\mathrm{LDR}(\astrosun)
\end{equation}

\begin{figure}
\centering
\subfigure{
  \includegraphics[scale=0.4]{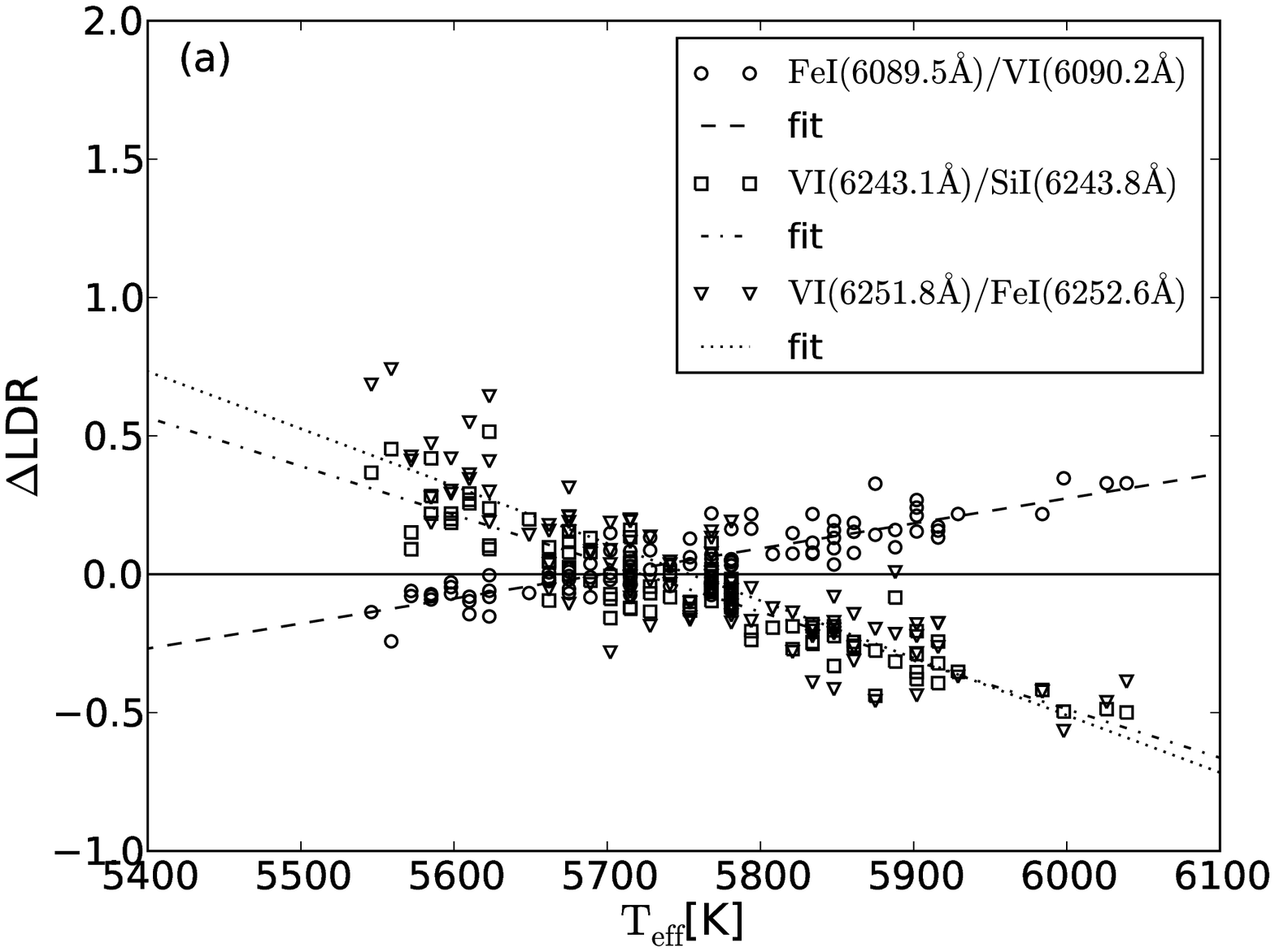}
\label{fig:LDRtemp}
}
\subfigure{
  \includegraphics[scale=0.4]{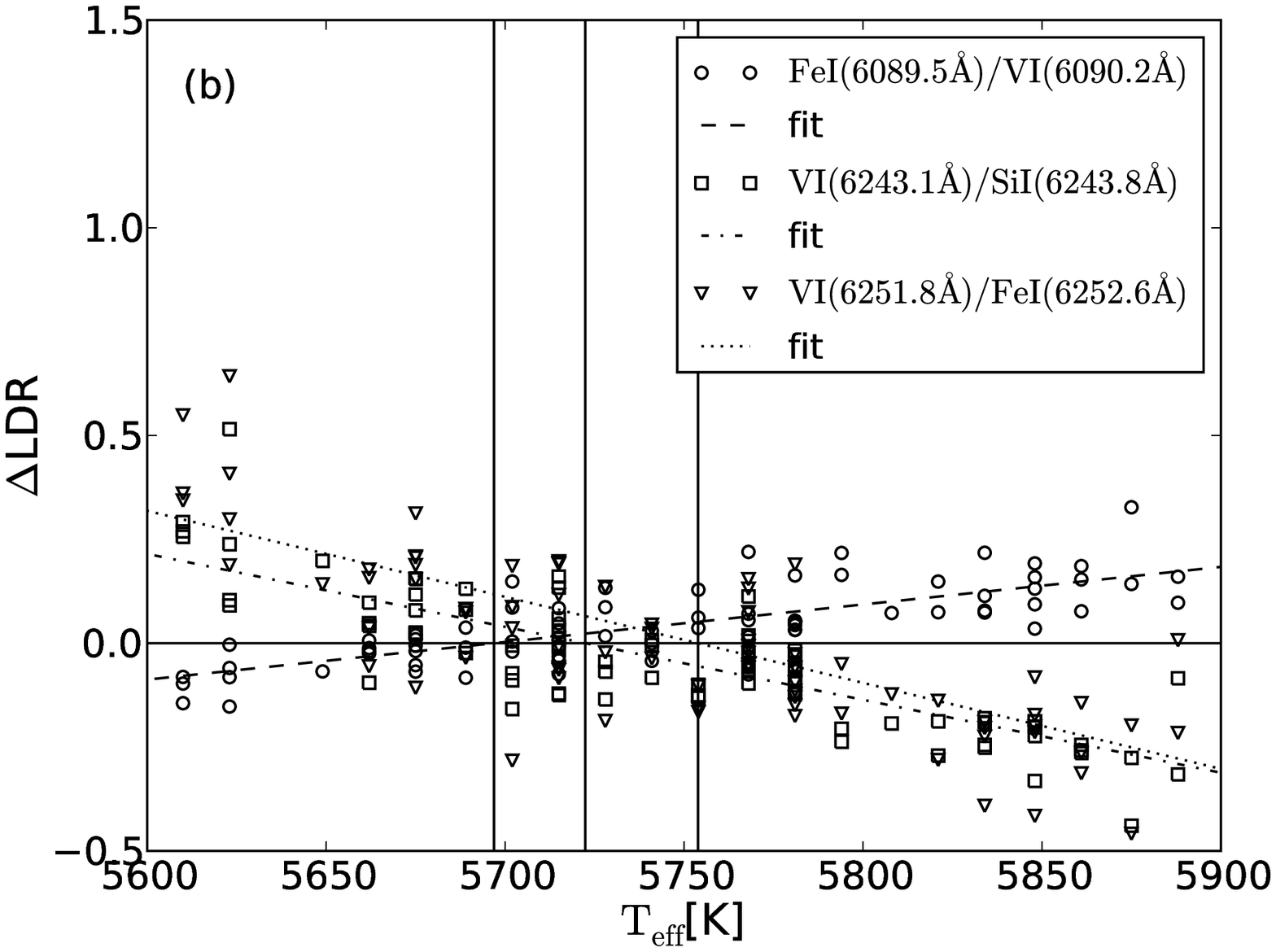}
\label{fig:LDRtemp2}
}
\caption[Optional caption for list of figures]{Panel (a): Line depth
    ratios for the three line pairs, relative to Ceres, shown as a function of
    GCS temperature and corresponding fits. Panel (b): Same line depth ratios, but
zoomed into the temperature range of 5600-5900\,K. The vertical black lines
guide the eye to the three crossing points with the zero line of the ratios.}
\label{fig:LDR}
\end{figure}

We show the three ratios as functions of the (GCS-III) temperature of the
sample stars in Fig.\,\ref{fig:LDR}. For each ratio there is a clear
temperature dependence, with a scatter that is consistent with the scatter in
the GCS temperatures.  We found no measurable dependence of the line ratios
with metallicity, as expected \citep{b16}.

We select solar twins with this method by requiring that the median relative
LDR vanishes. Analysis of the LDRs in the stars with repeated spectra show that
the typical observational error in the LDR is $\sigma = 0.04$, which we adopt
for all stars, given they all have similar S/N. Adopting $2\sigma$ limits and
using all three line pairs, we confirm four of our previous twin candidates
 (see Table 4). Since this method is sensitive only to the
temperature of the stars, it does not suffice by itself to select solar twins,
but needs to be combined with other methods that assess the metallicity. For
this reason we used this method only on twins already selected via the other
methods, as an additional check on their temperature match to the Sun.

\begin{table}
 \centering
     \caption{List of solar twins, confirmed by method (iv)}
  \begin{tabular}{@{}lrrr@{}}
  \hline
   & LDR, using & LDR, using & LDR, using\\
   Name & $\mathrm{FeI}_{6090}/\mathrm{VI}_{6090}$ & $\mathrm{VI}_{6243}/\mathrm{SiI}_{6243}$ & $\mathrm{VI}_{6252}/\mathrm{FeI}_{6253}$\\
 \hline
HD\,138573 & $-0.08\pm0.04$ & $0.08\pm0.04$ & $-0.04\pm0.04$\\
HD\,146233 & $0.02\pm0.04$ & $0.02\pm0.04$ & $-0.01\pm0.04$\\
HD\,147513 & $0.05\pm0.04$ & $-0.06\pm0.04$ & $-0.06\pm0.04$\\
HD\,163441 & $-0.02\pm0.04$ & $-0.09\pm0.04$ & $0.09\pm0.04$\\
\hline
\end{tabular}
\end{table}

It is clear from Table 4 that HD\,146233 (18\,Sco) is the best match using
method (iv). This star turned up in all four methods, and is certainly our best
overall twin.

\subsection{Multiple spectra and accuracy of the various methods}

Thanks to the large selection in the ESO archive, we were able to test the
repeatability of our results from multiple spectra of the same target. The best
examples are HD\,146233 and HD\,147513, for which we have six spectra of each
target. In Table 5 we show the $\chi^{2}(\Delta \mathrm{EW}_{\mathrm{all}})$,
$\textless\Delta \mathrm{EW}\textgreater$, $\textless\Delta
\mathrm{LD}\textgreater$ and slopes for each spectrum. The spectra were taken
in different nights and different years from 2004-2007 and show good agreement
in our measurements. The errors for the varius measuremed quantities are :
$\sigma(\chi^{2}(\Delta \mathrm{EW}_{\mathrm{all}}))$=0.2,
$\sigma(\textless\Delta \mathrm{EW}_{\mathrm{all}}\textgreater)$=0.002,
$\sigma(\textless\Delta \mathrm{EW}_{\mathrm{FeI}}\textgreater)$=0.002,
$\sigma$(slope[($\Delta \mathrm{EW}_{\mathrm{FeI}}$)
vs. $\chi{\mathrm{exc}}$])=0.001, $\sigma(\textless\Delta
\mathrm{LD}_{\mathrm{Fe}}\textgreater)$=0.004 and $\sigma$(slope[($\Delta
\mathrm{LD}_{\mathrm{FeI}}$) vs. $\chi{\mathrm{exc}}$])=0.001. These error
estimates have been adopted throughout section 4, since all stars have spectra
of very similar S/N.

\begin{table*}
 \centering
     \caption{Comparison of multiple spectra of the same object. Column 1 shows
       the HD number, and column 2 the spectrum running number. Column 3 and
       4 show the results for each spectrum using method (i), columns 5 and 6
       for method (ii); and columns 7 and 8 for method (iii) (see section 4.6
       for discussion of the typical errors of each method as derived from
       these multiple spectra.)}
  \begin{tabular}{@{}lccccccc@{}}
  \hline
   Name & Spectrum & $\chi^{2}(\Delta \mathrm{EW}_{\mathrm{all}})$ 
        & $\textless\Delta \mathrm{EW}_{\mathrm{all}}\textgreater$
        & $\textless\Delta \mathrm{EW}_{\mathrm{FeI}}\textgreater$ 
        & slope[($\Delta \mathrm{EW}_{\mathrm{FeI}}$) vs. $\chi{\mathrm{exc}}$]
        & $\textless\Delta \mathrm{LD}_{\mathrm{FeI}}\textgreater$
        & slope[($\Delta \mathrm{LD}_{\mathrm{FeI}}$) vs. $\chi{\mathrm{exc}}$]\\
 \hline

HD\,146233 &1 & 0.3 & 0.011 & 0.006 & 0.006 & $-$0.057 & 0.009\\
HD\,146233 &2 & 0.5 & 0.013 & 0.010 & 0.008 & $-$0.050 & 0.009\\
HD\,146233 &3 & 1.0 & 0.016 & 0.015 & 0.005 & $-$0.039 & 0.010\\
HD\,146233 &4 & 0.4 & 0.017 & 0.011 & 0.006 & $-$0.043 & 0.010\\
HD\,146233 &5 & 2.7 & 0.022 & 0.001 & 0.010 & $-$0.025 & 0.003\\
HD\,146233 &6 & 0.2 & 0.006 & 0.000 & 0.010 & $-$0.056 & 0.013\\
 \hline
HD\,147513 &1 & 0.6 & $-$0.026 & $-$0.023 & 0.006 & $-$0.067 & 0.015\\
HD\,147513 &2 & 1.2 & $-$0.023 & $-$0.022 & 0.006 & $-$0.063 & 0.011\\
HD\,147513 &3 & 1.6 & $-$0.029 & $-$0.021 & 0.005 & $-$0.066 & 0.010\\
HD\,147513 &4 & 0.6 & $-$0.024 & $-$0.005 & 0.009 & $-$0.044 & 0.016\\
HD\,147513 &5 & 0.8 & $-$0.028 & $-$0.021 & 0.006 & $-$0.068 & 0.013\\
HD\,147513 &6 & 0.9 & $-$0.033 & $-$0.020 & 0.005 & $-$0.067 & 0.012\\
\hline
\end{tabular}
\end{table*}

\subsection{Final list of solar twins}

Combining all these approaches and looking at the best twins from all four,
HD\,146233 is confirmed to be the best twin. We found two other stars,
HD\,126525 and HD\,138573, which satisfied the criteria within the 2$\sigma$
errors in three of our approaches; although, admittedly, in the case of
HD126525 these errors are rather large, and in fact its parameters are the most
extreme (it is a significantly cooler and more metal poor than the Sun according
to GCS-III (see Table 6)). Four stars are a match in two criteria, and three
are a match in one criterion. All stars satisfying any of our criteria are
shown with their stellar parameters (from GCS-III) in Table 6. The new
solar twins are identified by being shown in bold face.

\stepcounter{footnote} 
\begin{table}
 \centering
     \caption{Our solar twins compared to the Sun. Note that the solar
       $(\mathrm{b}-\mathrm{y})$ are as estimated indirectly from Sun-like stars by
       \citet{b19}. New twins are shown in bold face.}
  \begin{tabular}{@{}lrrrrlc@{}}
  \hline
   Name & $(\mathrm{b}-\mathrm{y})$ & $\mathrm{M_{V}}$ & [Fe/H] & $\mathrm{T}_{\mathrm{eff}}$ (K) & selection\\
   &&& (GCS) & (GCS) & method\\
 \hline
       Sun      & 0.403 & 4.83 & 0.00 & 5777 &\\
\hline
    HD\,146233  & 0.404 & 4.79 & $-$0.02 & 5768 & i, ii, iii, iv\\
\hline
\bf{HD\,126525} & 0.426 & 4.94 & $-$0.19 & 5585 & i, ii, iii\\
    HD\,138573  & 0.413 & 4.83 & $-$0.10 & 5689 & i, ii, iv\\
\hline
    HD\,78660   & 0.409 & 4.75 & $-$0.09 & 5715 & i, ii\\
\bf{HD\,117860} & 0.393 & 4.75 & $-$0.08 & 5821& i, ii\\
    HD\,147513  & 0.397 & 4.84 & $-$0.13 & 5781& ii, iv\\
\bf{HD\,163441} & 0.412 & 4.69 & $-$0.09 & 5702 & ii, iv\\
\hline
\bf{HD\,97356}  & 0.405 & 4.69 & $-$0.06 & 5754 & i\\
\bf{HD\,142415} & 0.383 & 4.66 & $+$0.04 & 5916 & i\\
\bf{HD\,173071} & 0.386 & 4.54 & $-$0.04 & 5875 & ii\\
\hline
\end{tabular}
\end{table}

\begin{figure}
	\includegraphics[width=90mm]{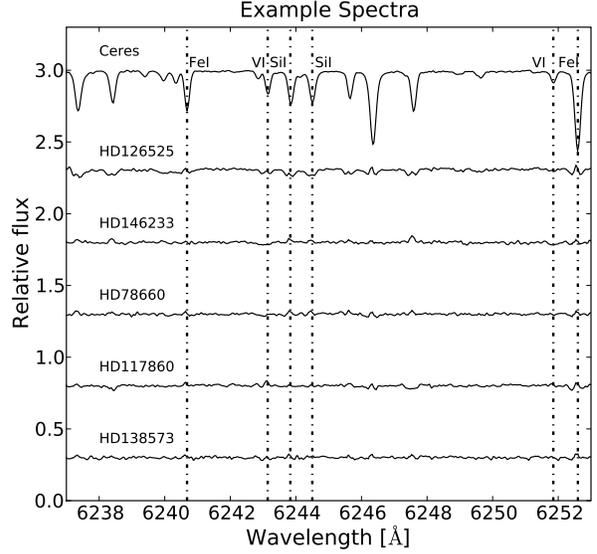}
	\caption{Comparison of the solar spectrum (Ceres) to some of our twins
          for an example region of the spectrum, including six of our lines
          (two FeI, two SiI and two VI). Each spectrum is shown as the residual
          relative to Ceres.}
	\label{examplespectrum}
\end{figure}

In Fig.\,\ref{examplespectrum} we illustrate for a small wavelength window
(6237 to 6253\,\AA), how similar our Ceres and solar twin spectra are, by
showing their residuals after subtracting the Ceres spectrum. While all the
stars are good matches, it is clear that small spectroscopic residuals
remain. Note that the vertical lines mark the positions of some of the lines
used for the comparison (FeI, SiI and VI lines) in this particular wavelength
window.

\begin{figure}
	\includegraphics[width=90mm]{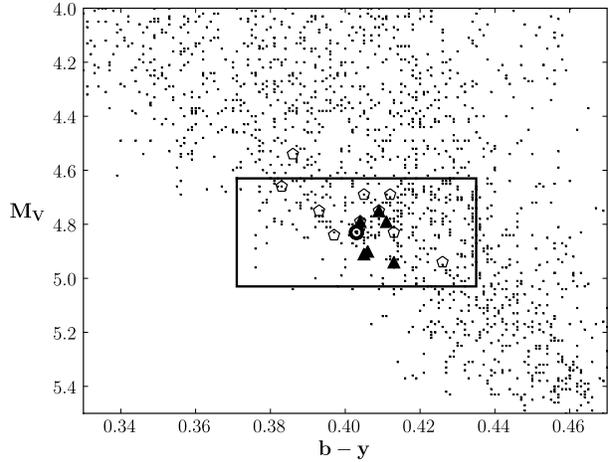}
	\caption{As Figure 1, but showing the locations of our solar twins as
          open pentagons. Dots show stars from the Nordstr\"om et al. (2004)
          catalogue (GCS-I) and the box is our original selection window.  The
          Sun is marked in the middle of the box, using the $(\mathrm{b}-\mathrm{y})$ colour
          estimated indirectly by \citet{b19} via solar analogues. The filled
          triangles show six of 11 solar twins by \citet{b32}, available in the GCS, for comparison.}
	\label{newsolartwins}
\end{figure}

Fig.\,\ref{newsolartwins} shows the location of these solar twins in the colour
magnitude diagram, compared to the original selection window.

\subsection{Offset of temperature and metallicity in the solar twins}

Table 6 shows that solar twins, proven to be a good match in photometric
(chosen in the original sample) as well as spectroscopic quantities (EW and LD
of spectral lines) can still be quite different from each other in their
individual photometric GCS-III temperatures and metallicities, by up to 200~K
and 0.2\,dex respectively. The scatter is even larger than the expected errors
on individual GCS entries: Nordstr\"om et~al.\ (2004) assume typical
uncertainties of order 0.1\,dex in metallicity and 0.01\,dex in $\log \mathrm{T}_{\mathrm{eff}}$
(or 135\,K, for solar temperatures). This suggests that even the differential
spectroscopic criteria used by us (and other authors) for the twin selection,
are still affected by metallicity-temperature degeneracy.

\begin{figure}
	\includegraphics[width=90mm]{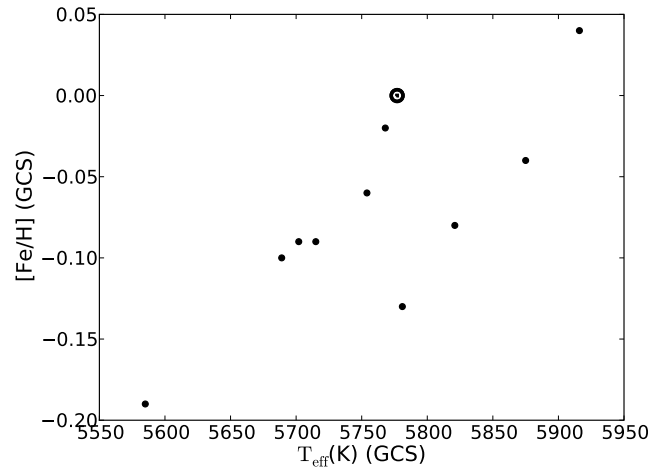}
	\caption{Effective temperatures and metallicities (from GCS-III) of our
          best solar twins (filled circles), compared to the solar values (dot
          with circle).  There is a trend to lower metallicities than solar in
          the sample of twins, while the mean temperature is close to the
          expected value of the Sun of 5777\,K.}
	\label{besttempmet}
\end{figure}

This is illustrated in Fig.\,\ref{besttempmet}, which shows our solar twins in
the GCS-III temperature--metallicity plane. Our twin sample shows a trend
towards lower metallicities than the Sun ($\mathrm{T}_{\mathrm{eff}}=$5777\,K
and [Fe/H]=0). In particular, nearly all the twins have sub-solar metallicity,
whereas the average twin temperature of $5760\pm20$\,K is quite close to the
solar $\mathrm{T}_{\mathrm{eff}}$.  However, especially if
temperature--metallicity degeneracies still affect the spectroscopic criteria,
the offset in [Fe/H] might simply reflect the fact that the peak of the
metallicity distribution function in the solar Neighbourhood, in the GCS scale,
is around $-0.15$\,dex; which naturally biases the selection toward metal--poor
siblings of the Sun. To acount for these potential biases, metallicity and
temperature for the full sample needs to be fit at the same time (i.e. section
5).

Offsets in the GCS scale, being about 100\,K too cool and 0.1\,dex too
metal--poor, have been suggested by \citet{b24}, by comparing their Infrared
Flux Method scale, to the average values of temperature and metallicity of 10
solar twins from Melendez et al. (2009).  For our twins, the average values
from Table 6 would suggest a metallicity offset of $-0.07\pm0.01$\,dex, but a
good temperature calibration within 20\,K of the Sun.

However, such offsets can hardly be assessed on the basis of only ten stars,
even though they are our best solar twins in the sample, considering that the
scatter in the twin properties, and the estimated errors on individual entries
in the GCS are of the same order or even larger than the offsets.  In the next
section we therefore devise an alternative method to estimate possible offsets in
the GCS scale with a broader approach, which uses our complete sample of
Sun--like stars.

\section{Probing the GCS temperature and metallicity scale}

As mentioned in the previous section, it is difficult to probe any offset in
the temperature or metallicity scales of the GCS on the basis of just 10
stars. In this section we use our whole sample of Sun--like stars to test the
calibration of the GCS scale, by introducing a new method that relies on the
systematic trends in the stellar spectra versus the reference solar/Ceres
spectrum.

\subsection{The degeneracy lines method}

The various quantities ($\textless\Delta \mathrm{EW}\textgreater$,
$\textless\Delta \mathrm{LD}\textgreater$ and corresponding slope[($\Delta
\mathrm{EW}_{\mathrm{FeI}}$) vs. $\chi_{\mathrm{exc}}$]) measured in our twin search depend both on
temperature and metallicity, as can be seen clearly from Fig \ref{fig:subfig1}
to \ref{fig:subfig4}.  Therefore we solved for this combined dependency by
applying 2-D least square fitting of a planar relation of the kind $a \,
[\mathrm{Fe/H}] + b \, {\mathrm{T}}_{\mathrm{eff}} + c$, using the GCS values
for [Fe/H] and $\mathrm{T}_{\mathrm{eff}}$.

We used 92 of our stars with temperatures over 5500\,K (the hottest being at
6039~K), as we found that below this threshold the dependencies no longer show
linearity; this behaviour, that linear fits are only suitable witihin a limited
$\mathrm{T}_{\mathrm{eff}}$ range, is also confirmed in our test with theoretical spectra in
Section~5.3. We have also experimented with narrower temperature ranges around
the solar value and verified that the fitting coefficients and the results are
quite stable to such changes.

We found the following 2-D planes for our data, with the errors in the fitting
coefficients being 5-10\%. Note, that the metallicity and temperatures in these
planes refer to the GCS values:

\begin{equation}
  \textless\Delta \mathrm{EW}_{\mathrm{all}}\textgreater =
  1.056[\mathrm{Fe/H}]-3.829\frac{\mathrm{T}_{\mathrm{eff}}-5777}{5777}+0.066
\end{equation}

\begin{equation}
  \textless\Delta \mathrm{EW}_{\mathrm{FeI}}\textgreater =
  0.856[\mathrm{Fe/H}]-3.769\frac{\mathrm{T}_{\mathrm{eff}}-5777}{5777}+0.045
\end{equation}

\begin{equation}
  \mathrm{slope}[(\Delta \mathrm{EW}_{\mathrm{FeI}})\ \mathrm{vs.}\ \chi_{\mathrm{exc}}]=
  0.256[\mathrm{Fe/H}]+0.027
\end{equation}

\begin{equation}
  \textless\Delta \mathrm{LD}_{\mathrm{FeI}}\textgreater =
  0.884[\mathrm{Fe/H}]-5.348\frac{\mathrm{T}_{\mathrm{eff}}-5777}{5777}-0.018
\end{equation}

\begin{equation}
  \mathrm{slope}[(\Delta \mathrm{LD}_{\mathrm{FeI}})\ \mathrm{vs.}\ \chi_{\mathrm{exc}}] =
  0.248[\mathrm{Fe/H}]+0.031
\end{equation}

\begin{figure*}
\centering
\subfigure{
  \includegraphics[scale=0.35]{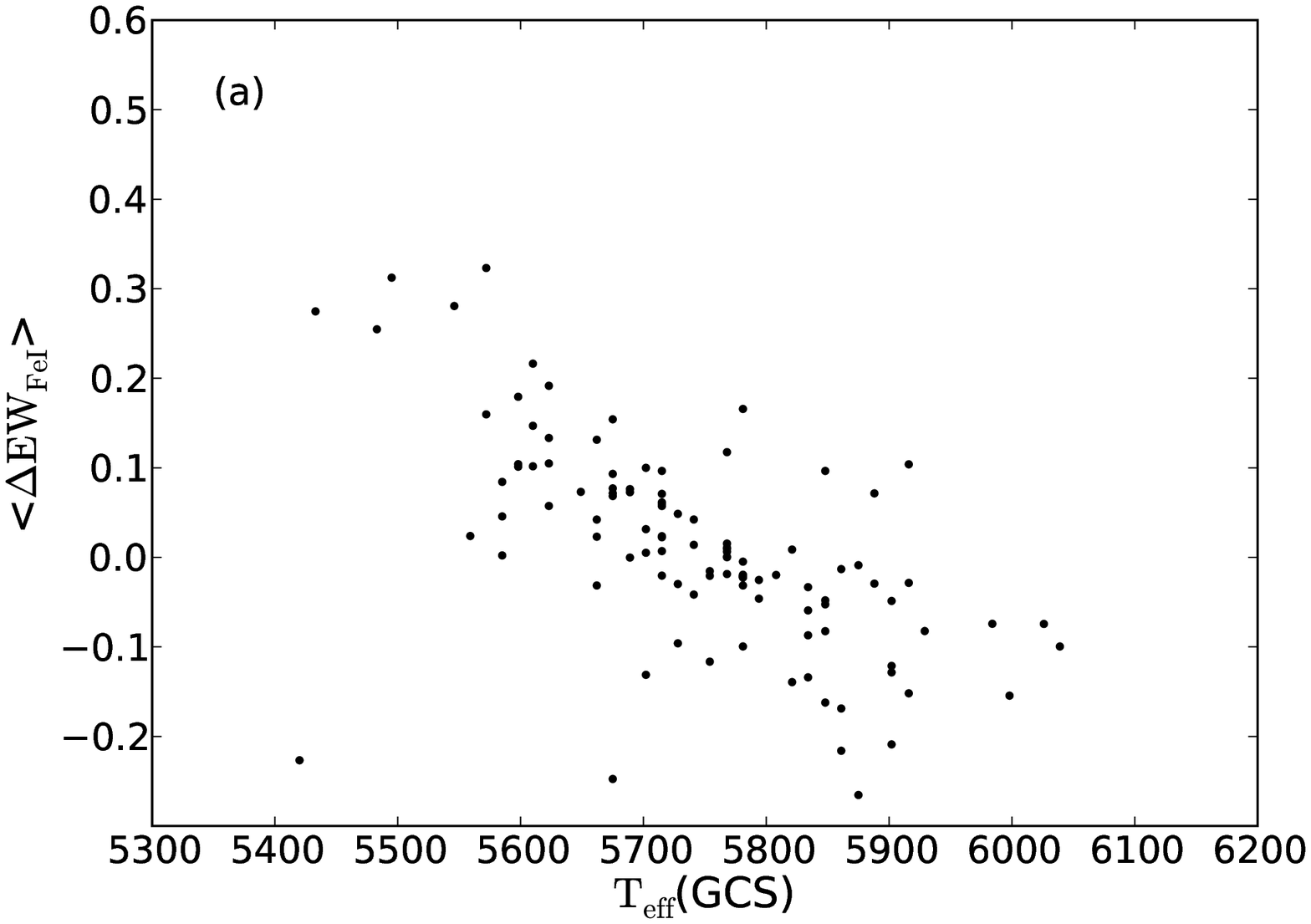}
\label{fig:subfig5}
}
\subfigure{
  \includegraphics[scale=0.35]{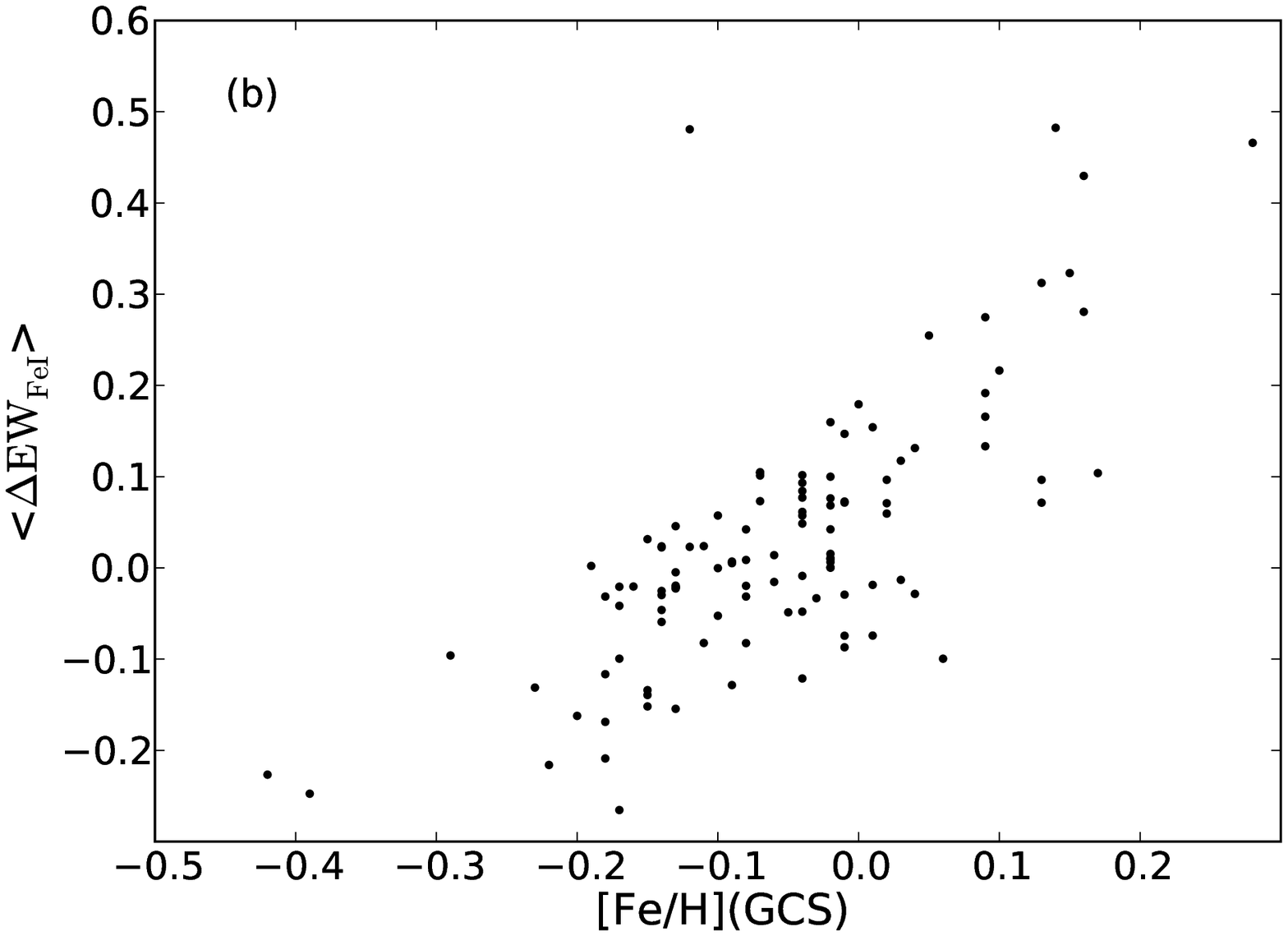}
\label{fig:subfig6}
}
\subfigure{
  \includegraphics[scale=0.35]{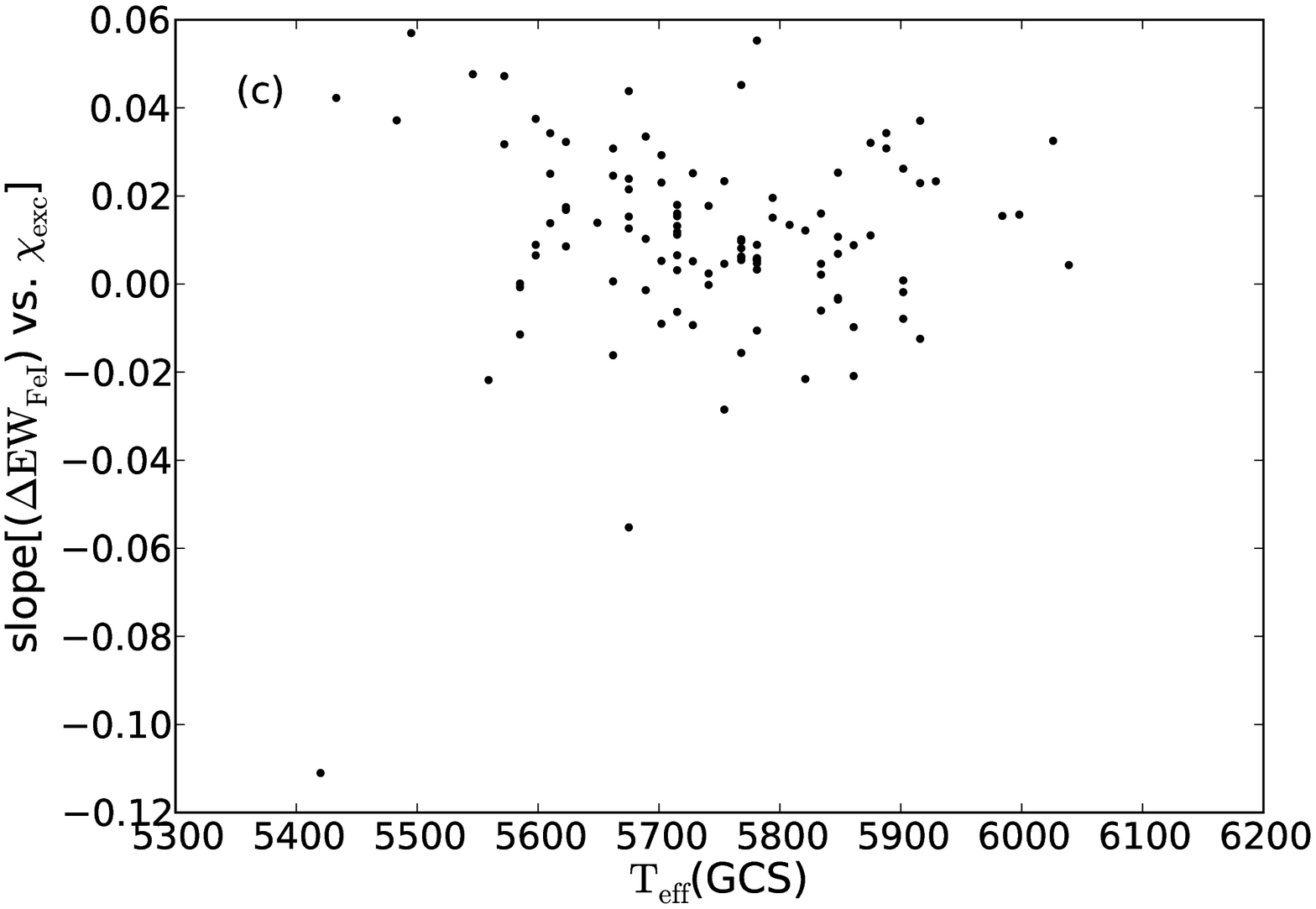}
\label{fig:subfig7}
}
\subfigure{
  \includegraphics[scale=0.35]{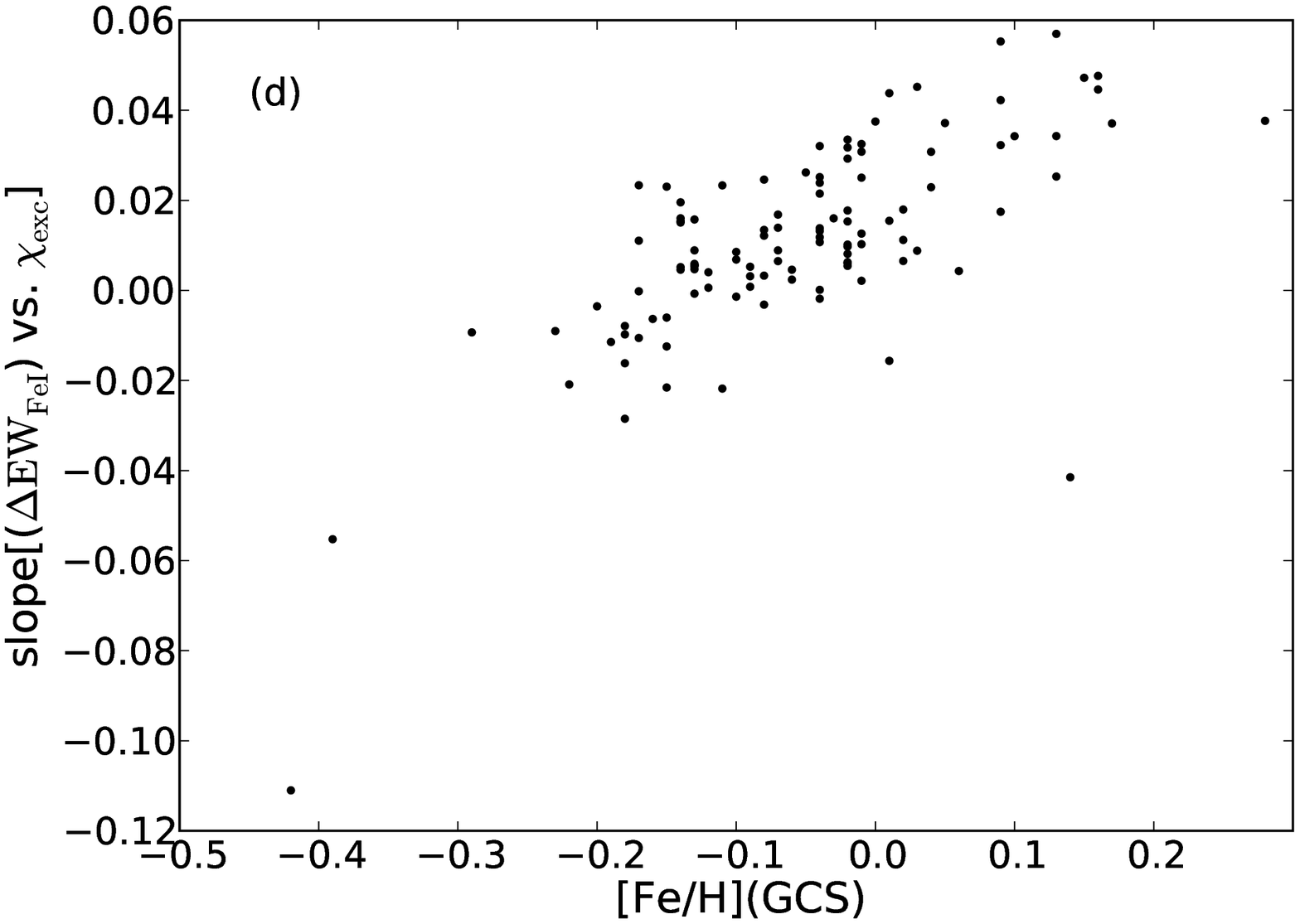}
\label{fig:subfig8}
}
\caption[Optional caption for list of figures]{For method (ii) (section 4.3),
  we show in panel (a) and (b) the dependancies of $\textless\Delta
  \mathrm{EW}_{\mathrm{FeI}}\textgreater$ on temperature and metallicity. In
  Panel (c) and (d), the same for slope[($\Delta \mathrm{EW}_{\mathrm{FeI}}$)
  vs. $\chi_{\mathrm{exc}}$].  We find no correlation of the slope with
  temperature. Planar fits of the form $a \, [\mathrm{Fe/H}] + b \,
  {\mathrm{T}}_{\mathrm{eff}} + c$ to the measured quantities allow us to solve
  for effective temperature and metallicity of the Sun in the GCS-III (section
  5).}
\label{fig:dependancies}
\end{figure*}

For Eq. 8 and 10 we found empirically that a metallicity dependence suffices to
describe the trend, and adding a temperature dependence does not significantly
improve the fits (see Fig.\,\ref{fig:dependancies}).

For the Sun, all the quantities on the left hand sides vanish; the fact that
the corresponding fitted relations on the right hand sides of Eq. 6-10 do not
(as the intercepts are non-zero) already hints to the existence of offsets in
the GCS-III temperatures and metallicities to the true solar values. Thus we
set the LHS to zero to measure this offset in the temperature and metallicity
on the GCS-III scale.

\begin{equation}
  [\mathrm{Fe/H}]_{\textless\Delta \mathrm{EW}_{\mathrm{all}}\textgreater} =
  3.625\frac{\mathrm{T}_{\mathrm{eff}}-5777}{5777}-0.062
\end{equation}

\begin{equation}
  [\mathrm{Fe/H}]_{\textless\Delta \mathrm{EW}_{\mathrm{FeI}}\textgreater} =
  4.406\frac{\mathrm{T}_{\mathrm{eff}}-5777}{5777}-0.053
\end{equation}

\begin{equation}
  [\mathrm{Fe/H}]_{\mathrm{slope}[(\Delta \mathrm{EW}_{\mathrm{FeI}})\
    \mathrm{vs.}\ \chi_{\mathrm{exc}}]} = -0.106
\end{equation}

\begin{equation}
  [\mathrm{Fe/H}]_{\textless\Delta \mathrm{LD}_{\mathrm{FeI}}\textgreater} =
  6.052\frac{\mathrm{T}_{\mathrm{eff}}-5777}{5777}+0.021
\end{equation}

\begin{equation}
  [\mathrm{Fe/H}]_{\mathrm{slope}[(\Delta \mathrm{LD}_{\mathrm{FeI}})\
    \mathrm{vs.}\ \chi_{\mathrm{exc}}]} = -0.124
\end{equation}

These relations correspond to (empirical) {\it degeneracy lines}, along which
the condition $\textless\Delta \mathrm{EW}\textgreater$$=0$, slope$=0$
etc.\ are respectively maintained. A star which may differ in metallicity and
temperature to the Sun, will have the same measured spectroscopic indices as
the Sun, along these lines of temperature-metallicity degeneracy. We have five
such relations, but they differ in their metal and temperature dependence,
allowing us to disentangle the degeneracies.

\begin{figure}
	\includegraphics[width=90mm]{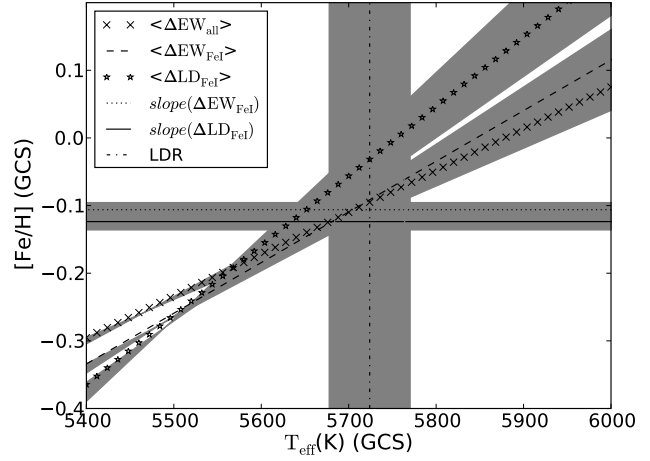}
	\caption{Derived relations for the various measured indices in the
          GCS-III temperature--metallicity plane. Grey areas mark the
          uncertainties in the fits. The solar values of the GCS scale are
          estimated to lie where the lines converge. We estimate this to lie at
          a metallicity of $\sim$ $-0.12$\,dex and a temperature of
          approximately 5680\,K, providing evidence that the GCS-III
          metallicity and temperature scales are shifted by about 0.1\,dex and
          100\,K from the Sun.}
	\label{finalplot}
\end{figure}

In Fig.\,\ref{finalplot} we plot these 5 relations in the GCS
temperature--metallicity plane, also including the solar temperature range we
get from the LDR measurements, see Fig.\,\ref{fig:LDRtemp2}.

As mentioned above, the solar values of metallicity and temperature lie where
all degeneracy lines would ideally cross, representing a simultaneous
zero--point for all of the relations above. There is no such single crossing
point, but we estimate the solar values to lie where the horizontal lines
(slopes) cross the inclined lines ($\Delta$). Thus the solar zero--point lies
around a GCS temperature of $\mathrm{T}_{\mathrm{eff}}=5680\pm40$\,K and a
metallicity of $[\mathrm{Fe/H}]=-0.12\pm0.02$. This corresponds to an offset in
the GCS-III scale of about $\Delta$T = $-$100\,K and $\Delta$[Fe/H] =
$-$0.1\,dex with respect to the true solar values, meaning the GCS scale seems
to be too cold and too metal poor, at least around the solar temperature. This
confirms the findings of \citet{b24}, who found similar offsets.

The offsets we quote are based on the various spectroscopic quantities relevant
for methods (i)-(iii) of Section 4. We note that method (iv), which is only
sensitive to temperature, yields a smaller but still non-negligible offset of
about $-$50\,K in the temperature scale of GCS-III (see Fig. 14).

\subsection{The solar $(\mathrm{b}-\mathrm{y})$ colour}

\begin{figure}
	\includegraphics[width=90mm]{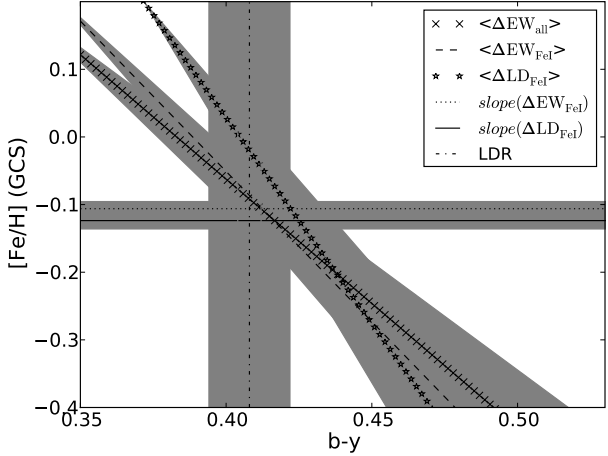}
	\caption{Degeneracy lines between metallicity and temperature for the
          quantities in the various methods in section 4, shown in the
          $(\mathrm{b}-\mathrm{y})$--metallicity plane; the solar values lie
          where the degeneracy relations converge. We estimate a solar
          $(\mathrm{b}-\mathrm{y})$ of $0.414\pm0.007$ from this plot, in good
          agreement with the recent estimate by \citet{b23} of $0.411\pm0.002$.}
	\label{finalplotby}
\end{figure}

We applied an analogous procedure to the previous section, using the
$(\mathrm{b}-\mathrm{y})$ colour instead of temperature to get an estimate of
this colour for the Sun. The results are shown in Fig.\,\ref{finalplotby}. We
estimate a solar $(\mathrm{b}-\mathrm{y})$ colour of $0.414\pm0.007$, which on
is in good agreement with the very precise value by \citet{b23} of
$0.411\pm0.002$, but is in mild tension with our initially assumed value
  in selecting the sample of solar twin candidates (i.e. $0.403 \pm 0.013$;
  Holmberg et al (2006)). The initial colour selection window was much wider
  than this small change to the colour, so this is very unlikely to have biased
  the sample.

This procedure yields another estimate for the metallicity of a solar twin on
the GCS-III scale: we get $\mathrm{[Fe/H}] = -0.11\pm0.01$. This is in good
agreement with the result in the previous section.

\subsection{Testing the method : internal precision and synthetic spectra}

The method of solving for the crossing points of differing temperature and
metallicity degeneracy relations is novel, so we have performed a number of
tests to validate that it returns the required results.

Our first test was to use the method on randomly chosen reference stars in the
data set, rather than the Ceres spectrum. We ran the same routines and used the
same approach to determine the crossing point of the degeneracy lines as
described in section 5.1. This tests whether the temperature and metallicities
of the reference star can be recovered within the GCS scale. We found that our
method indeed recovers the GCS temperatures and metallicities very well:
comparing the recovered temperatures and metallicities to GCS values of the
reference star, we get typical offsets of $\Delta$T = 24\,K with a scatter of
50\,K, and $\Delta$[Fe/H] = 0.00 with a scatter of 0.06\,dex. These are small
compared to the offsets we find between the Ceres spectrum and the GCS scale,
of $\approx 100$\,K and 0.1\,dex.  In no case do we find a {\it combined} error
on the recovered temperature and metallicity as large as what we find for
Ceres; hence we are confident that the GCS offsets are real. Table 7 summarises
some results from this test.

\begin{table}
 \centering
 \caption{Tests of the degeneracy relations method. 
     Column 1 shows the reference star, and columns 2 and 3 its GCS
     temperatures and metallicities.  Columns 4 and 5 show the values recovered
     from our degeneracy line fitting, by comparing the reference star to the
     rest of the sample stars.}
  \begin{tabular}{@{}lcrcr@{}}
  \hline
   Name & GCS & GCS  & recovered  & recovered\\
   &$T_{\mathrm{eff}}$&[Fe/H]&$T_{\mathrm{eff}}$& [Fe/H]\\
   \hline
  HD\,12264  & 5728 & $-$0.14 & 5733 & $-$0.11\\
  HD\,78660  & 5715 & $-$0.09 & 5608 & $-$0.10\\
  HD\,126525 & 5585 & $-$0.19 & 5538 & $-$0.19\\
  HD\,138573 & 5689 & $-$0.10 & 5616 & $-$0.15\\
  HD\,155968 & 5662 &    0.04 & 5683 &    0.07\\
  HD\,222669 & 5834 & $-$0.03 & 5846 & $-$0.05\\ 
 \hline
\end{tabular}
\end{table}

As a final check on our method in section 5.1, we applied it to a set of
synthetic spectra of Sun-like stars. The spectra were taken from the ``Grids of
ATLAS9 Model Atmospheres and MOOG Synthetic Spectra'' computed by \citet{b34},
and available from the VizieR On-line Data Catalog (Catalog code VI/134) at
SIMBAD (http://simbad.u-strasbg.fr/simbad/). The spectra have been computed
using MOOG \citep{b35}, and cover the grid $3500$\,K$ \le \mathrm{T}_{\mathrm{eff}} \le
8000$\,K, $0.0 \le \mathrm{log\,g} \le 5.0$, $-4.0 \le $[M/H]$ \le 0.0$, and
$-0.8 \le [\alpha$/Fe] $\le +1.2$, in the wavelength range 6300\,\AA\ to
9100\,\AA, with a resolution of 20\,m\AA, slightly better than that in our data
(30\,m\AA). Around the Sun, spectra are available in the library at [Fe/H] $=
-0.2, -0.1$ and 0.0 and ($\mathrm{T}_{\mathrm{eff}}$ = 5500, 5600, 5800, 6000 and
6200\,K). We selected spectra for which $\mathrm{log\,g} = 4.5$ as
representative of main sequence stars, and [$\alpha$/Fe] = 0.0, as alpha
enhancement in this metallicity range is mild, and our lines are mainly due to
Fe. This resulted in 15 synthetic spectra very similar to our observational
spectra in their basic stellar properties.  Our line list contains 95 lines in
the range 5044\,\AA\ to 7836\,\AA, but most of the lines are blueward of the
lower wavelength cutoff of 6300\,\AA\ of the theoretical spectra, so that we
only had 27 lines left after taking this into account. We measured equivalent
widths for these 27 lines in the synthetic spectra using TWOSPEC, using the
spectrum at $\mathrm{T}_{\mathrm{eff}} = 5800$\,K and [Fe/H] $= 0.0$ as our
reference spectrum, and applied our basic method to the results. The library
spectra are noisefree, and we sampled them with a signal to noise
(conservatively) of 100:1.

Doing this we get
5814\,K and 0.012\,dex for the reference spectrum, compared to the input values
of 5800\,K and 0.000\,dex. Fitting to just the temperature range 5600 to
6000\,K, we recover a temperature and metallicity of 5803\,K and 0.004\,dex, so
the choice of temperature and metallicity range have only a small effect on the
solutions. Interestingly, sampling the theoretical spectra with a signal to
noise of only 20:1, we recover an effective temperature and metallicity for the
reference spectrum of 5800\,K and $-$0.01\,dex, demonstrating that the method
is robust to spectra of rather poor quality.

Some time after completing the MOOG study above, we found a much more extensive
library of spectra, both in metallicity and temperature coverage, as well as
spectral coverage, in the Pollux spectral library \citep{b38}. The
spectra also have 20 m\AA\ resolution and cover our
full wavelength range of analysed lines. 

Sixty grid points around a "Solar" model at $\mathrm{T}_{\mathrm{eff}}$ = 5750
K, logg = 4.5 and [Fe/H] = 0.0 were selected from the library, at grid points
of $\mathrm{T}_{\mathrm{eff}}$ 5500, 5750, 6000 and 6250 K, and metallicities
of $-0.5$, $-0.25$, 0.0, 0.25 and 0.5. We applied the same method as used on
the MOOG spectra but now with a much larger line list because of the full
spectral coverage, and inserted similar noise as in our real spectra. We fully
confirm our findings with the MOOG spectra, that the method recovers the
correct temperature (within 30 K) and metallicity (within 0.02 dex) of the
reference star.

These studies of synthetic spectra offer substantial support for our basic
methodology, showing we can recover the temperature and metallicity of
reference stars from a sample of stars with similar metallicities and
temperatures. We could improve the analysis substantially by using a much finer
grid of temperatures, spaced by 50 K instead of 100 to 200\,K. This would
involve computing dedicated spectra, rather than drawing from a
pre-computed library, and we leave this to future work.

\section{Summary and conclusions}

In this paper we use high resolution optical spectroscopy to search for solar
twins in the Geneva Copenhagen Survey, by applying various methods adopted from
recent literature. We have shown that there is no unique way to search for a
solar twin and that it is necessary to include photometric as well as
spectroscopic selection criteria to really determine which stars are solar
twins.

We confirm HD\,146233 (18\,Sco) as the best twin in our list, being selected by
all 4 spectroscopic methods used; HD\,126525 and HD\,138573 are second best,
being selected in 3 out of 4 methods; 6 out of our 10 twins are new additions
to the literature; HD\,117860, HD\,97356, HD\,142415, HD\,163441, HD\,173071
and HD\,126525.

We use our entire sample to probe for offsets in the temperature and
metallicity scale in the GCS for Sun--like stars, introducing a new method (degeneracy lines method)
which disentangles the differing metallicity and temperature degeneracies 
in the measured indices for our stars.

We estimate that, for Sun--like stars, the GCS-III scale is offset by
$(-0.12\pm0.02)$\,dex and $(-97\pm35)$\,K respectively -- i.e. we find it is a
little too metal poor and cool. This result is in good agreement
with similar offsets claimed in recent literature, based on solar twins:
\citet{b23} find the GCS values to be $\Delta$T = 48\,K too cool and
$\Delta$[Fe/H] = 0.09 too metal poor. \citet{b24} finds offsets of about
$-$100\,K and $-$0.1\,dex, respectively.

Our new method has been successfully tested both internally in GCS
(i.e.\ recovering the metallicity and temperature of random reference stars,
which replaced the Sun/Ceres for the sake of the test) and on theoretical
spectra.  We are currently applying a similar degeneracy lines method to the
very high quality High Accuracy Radial velocity Planet Searcher (HARPS) archive spectra, 
using both neutral and ionised species for more elements than just
Fe. Early results confirm the offsets found here and will be discussed in a
forthcoming paper (Datson et al., in preparation).

Despite the agreement of our results with other studies for an offset in the
temperature and metallicity scales of GCS for Sun--like stars, we
point out that recent measurements of the temperatures of Sun-like stars via
interferometry with the Center for High Angular Resolution Astronomy (CHARA) array 
instrument \citep{b37} show, for over
a dozen stars with the most secure angular diameters ($>$1~mas) {\it excellent
  agreement} with GCS temperatures (Holmberg, private comm.). 
This contrasts with the conclusions drawn by us and other authors, based on
solar twins and Sun-like stars. We
leave the discussion of this intriguing result to future work.

The offsets we find in the GCS would imply the solar $(\mathrm{b}-\mathrm{y})$
colour to be $(\mathrm{b}-\mathrm{y}) = 0.414\pm0.007$, also determined via our
degeneracy--lines approach.  This colour is redder than the
$(\mathrm{b}-\mathrm{y})=0.403\pm0.013$ found earlier by our group (Holmberg et
al. 2006) but very close to the recent result of Melendez et al.\ (2010), based
on solar twins.

One of our best solar twins, HD\,126525, has a temperature and metallicity,
that are so offset from solar (5585 K and $-0.19$ dex, see Table~6), that even
the proposed corrections to the GCS scale still leave it in tension with the
solar values. We cannot rule out that the twin selection methods used here are
still affected by systematics; a more detailed study of the individual spectra
and of metallicity-temperature degeneracy issues is currently underway.

Three of our twins are known to host an exoplanet: HD\,142415, HD\,147513
\citep{b30} and HD\,126525 \citep{b31}. There have been no confirmed 
detections of exoplanets for the other seven twins so far, which will
hopefully be included as targets for future planet searches.

\section*{Acknowledgments}

We would like to thank Johan Holmberg for insightful and constructive criticism
on our work; Ivan Ram{\'{\i}}rez and Jorge Mel{\'e}ndez for useful discussions;
A.~Mueller, B.~Conn, A.~Ederoclite and the anonymous ESO staff observers, for
the data acquisition at the telescope. We would also like to thank the referee for
thorough reading and very constructive criticisms. This research has made use of the SIMBAD
database, operated at CDS, Strasbourg, France.

This study was financed by the Academy of Finland (grant nr.~130951 and 218317)
and the Beckwith Trust. We thank the University of Sydney and Swinburne
University, where part of this work was carried out.

\bsp

\label{lastpage}

\end{document}